\documentclass[aps,prd,twocolumn,10pt,groupedaddress]{revtex4-1}
\usepackage{amssymb}
\usepackage{graphicx}
\usepackage{amsmath}
\usepackage{hyperref}

\begin{document}

\title{Higgs inflation in Gauss-Bonnet braneworld}
\author{Rong-Gen Cai}
\email{cairg@itp.ac.cn}
\author{Zong-Kuan Guo}
\email{guozk@itp.ac.cn}
\author{Shao-Jiang Wang}
\email{schwang@itp.ac.cn}
\affiliation{State Key Laboratory of Theoretical Physics, Institute of Theoretical Physics, Chinese Academy of Sciences, Beijing 100190, China}
\date{\today}

\begin{abstract}
The measured masses of the Higgs boson and top quark indicate that the effective potential of the standard model either develops an unstable electroweak vacuum or stands stable all the way up to the Planck scale. In the latter case in which the top quark mass is about $2\sigma$ below its present central value, the Higgs boson can be the inflaton with the help of a large nonminimal coupling to curvature in four dimensions. We propose a scenario in which the Higgs boson can be the inflaton in a five-dimensional Gauss-Bonnet braneworld model to solve both the unitarity and stability problems which usually plague Higgs inflation. We find that in order for Higgs inflation to happen successfully in the Gauss-Bonnet regime, the extra dimension scale must appear roughly in the range between the TeV scale and the instability scale of standard model. At the tree level, our model can give rise to a naturally small nonminimal coupling $\xi\sim\mathcal{O}(1)$ for the Higgs quartic coupling $\lambda\sim\mathcal{O}(0.1)$ if the extra dimension scale lies at the TeV scale. At the loop level, the inflationary predictions at the tree level are preserved. Our model can be confronted with future experiments and observations from both particle physics and cosmology.
\end{abstract}
\maketitle

\section{Introduction}\label{sec:1}

The recently released \emph{Planck} 2015 data \cite{Ade:2015lrj} provide growing evidences that our observable universe has experienced an inflationary era, stretching the primordial quantum fluctuations to the cosmic size, leaving distinct imprints on the Cosmic Microwave Background Radiation and seeding the formations of cosmic structures. The current favoured inflationary scenarios \cite{Martin:2015dha} are those single-field slow-roll inflationary models, where the scalar field plays the role of inflaton. Despite the phenomenological success of inflation, there is growing theoretical interest to connect inflation with the low-energy particle physics, among which Higgs inflation is the most attractive model due to its minimality.

Higgs inflation \cite{Bezrukov:2007ep} makes use of a nonminimal coupling $\xi$ of the standard model (SM) Higgs boson to four-dimensional Einstein gravity. At the high-energy scale, the Higgs boson is decoupled from SM and slowly rolls down an exponential plateaulike potential in the Einstein frame. The \emph{Planck} normalization requires a large nonminimal coupling $\xi\simeq5\times10^4\sqrt{\lambda}\simeq1.8\times10^4$ for tree-level estimation of the Higgs quartic coupling $\lambda\simeq m_h^2/2v^2\simeq0.13$ from the Higgs mass $m_h\simeq125$ GeV and vacuum expectation value (VEV) $v\simeq246$ GeV. At intermediate energy scale where preheating\cite{GarciaBellido:2008ab}/reheating \cite{Bezrukov:2008ut} came to play, the Higgs boson oscillates along a quadratic potential and decayed into SM particles. At the low-energy scale, the potential is transited into the usual SM quartic potential. The cosmological predictions of Higgs inflation can fit the \emph{Planck} 2015 data well and exhibit insensitivity to its reheating processes \cite{Cai:2015soa}. However, there are two major problems plaguing Higgs inflation: the unitarity problem and the stability problem.

The stability problem \cite{Allison:2013uaa,Salvio:2013rja} states that: for a successful Higgs inflation, the top quark mass is required to be about $2\sigma$ below its present central value for the measured Higgs mass. The stability problem of Higgs inflation shows that there is a potential tension between constraints from particle physics and those from cosmology. To stabilize the SM electroweak (EW) vacuum in Higgs inflation, one either introduces new particles thresholds such as scalar fields \cite{EliasMiro:2012ay,Gabrielli:2013hma,Haba:2014zda,Barbon:2015fla}, fermion fields \cite{Gogoladze:2008ak,He:2012ub,Okada:2015zfa} and vector field \cite{He:2014ora}, or invokes new physics such as asymptotically safe Higgs inflation \cite{Shaposhnikov:2009pv,Cai:2012qi,Cai:2013caa,Xianyu:2014eba}. It is worth noting that Higgs inflation can also be realized \cite{Bezrukov:2014ipa} in the case of a metastable EW vacuum if one takes into account the unknown finite parts of counterterms and finite temperature corrections to the effective potential.

The unitarity problem \cite{Burgess:2009ea,Barbon:2009ya,Lerner:2009na,Hertzberg:2010dc,Burgess:2010zq,Bezrukov:2010jz,Lerner:2011it,Xianyu:2013rya,Ren:2014sya} states that: the tree-level analysis is already invalid even before Higgs inflation can take place at the scale $M_P/\sqrt{\xi}$ due to unitarity violation at the scale $M_P/\xi$ by naive power-counting. Restoring unitarity above $M_P/\xi$ introduces either new particles or new interactions, both of which might spoil the flatness of the inflationary potential in an uncontrollable manner. There are three ways to address the unitarity problem:
First, introducing new interactions such as new Higgs inflation \cite{Germani:2010gm}, unitary Higgs inflation \cite{Lerner:2010mq}, the Higgs $\sigma$ model \cite{Giudice:2010ka}, and its variant \cite{Barbon:2015fla}. However, there is no guarantee \cite{Lerner:2011it} whether the quantum corrections of these new interactions are under control.
Second, recognizing the background dependent cutoff \cite{Bezrukov:2010jz,Xianyu:2013rya,Ren:2014sya} above which the strong dynamics should enter to restore unitarity. However, there is also no guarantee \cite{Lerner:2011it} whether the strong dynamics would call for new physics.
Third, fine-tuning the Higgs mass and top quark mass to achieve an extremely small Higgs quartic coupling around the Planck scale as in the case of critical Higgs inflation \cite{Hamada:2013mya,Cook:2014dga,Hamada:2014iga,Bezrukov:2014bra,Hamada:2014wna}. However, an unnaturally small $\lambda$ requires the top quark mass being about $2\sigma$ below its present central value, and $\xi$ can only be made of $\mathcal{O}(1)$ if one allows a large $r\gtrsim0.1$ in direct conflict with \emph{Planck} 2015 $\mathrm{TT,TE,EE+lowP}$ constraints \cite{Ade:2015lrj}.
We report in this paper an alternative: extra dimensions.

The idea of extra dimensions stemmed from the attempt by Kaluza and Klein to unify the gravitational and electromagnetic interactions. Although the idea failed, the formalism survived. Later it was found that string theory can only be defined consistently in higher dimensions while the compactification scale is too high to be tested experimentally. However, the large extra dimension scenarios renewed the interest of extra dimensions in Arkani-Hamed, Dimopoulos and Dvali (ADD) model \cite{ArkaniHamed:1998rs,Antoniadis:1998ig} and Randall and Sundrum (RS) models \cite{Randall:1999ee,Randall:1999vf}, and opened new door to tackle those profound mysteries in particle physics and cosmology. In five dimensions, it is natural to include the Gauss-Bonnet term for four reasons: First \cite{Lovelock:1971yv}, it presents an unique combination of a second order symmetric and divergence-free tensor that can lead to second order field equations in bulk metric components. Second \cite{Gross:1986mw}, it arises in the heterotic string theory as next-to-leading order corrections with the Gauss-Bonnet coupling identified with Regge slope. Third \cite{Zwiebach:1985uq}, it leads to ghost-free nontrivial gravitational self-interactions for dimensions higher than four. Fourth \cite{Kim:2000ym,Kim:2000pz,Mavromatos:2000az,Neupane:2000wt,Neupane:2001st,Meissner:2000dy,Meissner:2001xg}, the zero mode of graviton is localized on the brane at low energy with only two independent degrees of freedom corresponding to the usual four-dimensional graviton. As a result, there are extensive studies on the Gauss-Bonnet braneworld scenario.

In this paper, we realize Higgs inflation in the five-dimensional Gauss-Bonnet braneworld cosmology. We find that, for Higgs inflation happened in the Gauss-Bonnet regime, the combined parameter $\lambda/\xi^2$ could increase many orders of magnitudes with decreasing energy scale of the extra dimension, and the extra dimension scale must appear roughly in the range between the TeV scale and the SM instability scale. For the extra dimension scale near the TeV scale, the nonminimal coupling can be made of $\xi\sim\mathcal{O}(1)$ for the Higgs quartic coupling $\lambda\sim\mathcal{O}(0.1)$ with tensor-to-scalar ratio $r\sim10^{-12}$ safely inside \emph{Planck} 2015 $\mathrm{TT,TE,EE+lowP}$ bound $r\lesssim0.1$. The prediction of scalar spectral index $0.960\lesssim n_s\lesssim0.968$ and its running $-0.0008\lesssim\alpha_s\lesssim-0.0005$ remains almost the same as in the four-dimensional case for all possible extra dimension scale. Furthermore, the inflationary predictions are preserved beyond tree-level analysis.

This paper is organized as follows. In Sec. \ref{sec:2}, We review the general formalism of the five-dimensional Gauss-Bonnet braneworld scenario. In Sec. \ref{sec:3}, we propose Higgs inflation in the five-dimensional Gauss-Bonnet braneworld model. In Sec. \ref{sec:4}, the tree-level results are summarized. In Sec. \ref{sec:5}, we go beyond tree-level analysis. The last section is devoted to conclusions.

\section{Gauss-Bonnet Braneworld Cosmology}\label{sec:2}

We briefly review in this section the general formalism of the five-dimensional Gauss-Bonnet braneworld scenario.

The total action of the Guass-Bonnet braneworld model reads (we neglect possible boundary terms)
\begin{eqnarray}\label{eq:action}
\nonumber S_5&=&\frac{1}{2\kappa_5^2}\int_{\mathrm{AdS_5}}\mathrm{d}^5x\sqrt{-g_5}\left[-2\Lambda_5+R_5\right.\\
\nonumber     &&\left.+\alpha\left(R_5^2-4R_{ab}^{(5)}R_{(5)}^{ab}+R_{abcd}^{(5)}R_{(5)}^{abcd}\right)\right]\\
              &&+\int_{\mathrm{brane}}\mathrm{d}^4x\sqrt{-g_4}\left(-m_{\sigma}^4+\mathcal{L}_{\mathrm{matter}}\right),
\end{eqnarray}
which contains a five-dimensional anti-de Sitter (AdS) bulk with a negative cosmological constant $\Lambda_5$ and a four-dimensional Friedmann-Robertson-Walker (FRW) brane with a positive tension $m_{\sigma}^4$. The confined matter field with Lagrangian density $\mathcal{L}_{\mathrm{matter}}$ can be approximated as the perfect fluid by assumption. The Gauss-Bonnet term is weighted by $\alpha$, which should be positive in the view of stringy generalisation of general relativity for Einstein-Gauss-Bonnet gravity. We will see that the Planck scale $M_4(\kappa_4^2=1/M_4^2=8\pi G_4)=2.435\times10^{18}\mathrm{GeV}$ on the four-dimensional FRW brane can be derived from the more fundamental Planck scale $M_5(\kappa_5^2=1/M_5^3=8\pi G_5)$ in the five-dimensional AdS bulk.

The field equation and junction equation of the action (\ref{eq:action}) admit a FRW brane solution
\begin{equation}
\mathrm{d}s_4^2=-\mathrm{d}t^2+a(t)^2\gamma_{ij}\mathrm{d}x^i\mathrm{d}x^j,
\end{equation}
which can be induced from the AdS bulk metric,
\begin{equation}
\mathrm{d}s_5^2=-f(a)\mathrm{d}\tau^2+\frac{\mathrm{d}a^2}{f(a)}+a^2\gamma_{ij}\mathrm{d}x^i\mathrm{d}x^j,
\end{equation}
by requiring
\begin{equation}
-f(a)\dot{\tau}(t)^2+\frac{\dot{a}(t)^2}{f(a)}=-1,
\end{equation}
with respect to the embedding coordinates $\tau(t)$ and $a(t)$. Therefore, the scale factor $a(t)$ on the brane can be interpreted as the motion of brane $a(\tau)$ in the bulk. Here $\gamma_{ij}$ describes a maximally symmetric 3-hypersurface with spatial curvature constant $k_3=0,\pm1$
and $f(a)$ can be solved for pure AdS spacetime \cite{Boulware:1985wk,Cai:2001dz} as
\begin{equation}
f(a)=k_3+a^2\mu^2,
\end{equation}
where $\mu$ has two branches for $\alpha>0$,
\begin{equation}\label{eq:mu}
\mu^2=\frac{1}{4\alpha}\left(1\pm\sqrt{1+\frac{4}{3}\alpha\Lambda_5}\right),
\end{equation}
and the negative branch of (\ref{eq:mu}),
\begin{equation}\label{eq:lambda}
\Lambda_5=-6\mu^2(1-2\alpha\mu^2),
\end{equation}
has the RS limits $\Lambda_5=-6\mu^2$ by taking $\alpha\rightarrow0$. $\mu$ is usually associated with bulk curvature scale $|R_5|\sim\mu^2$. Introducing a dimensionless Gauss-Bonnet coupling $\beta\equiv4\alpha\mu^2$, then the subdominated Gauss-Bonnet term $\alpha|R_5^2|\ll|R_5|$ requires $\beta\ll4$. The negative bulk cosmological constant $\Lambda_5<0$ requires $\beta<2$ from (\ref{eq:lambda}) and the negative branch $1-4\alpha\mu^2<0$ requires $\beta<1$ from (\ref{eq:mu}).

The modified FRW equation now reads \cite{Charmousis:2002rc,Maeda:2003vq}
\begin{equation}\label{eq:FRW1}
\kappa_5^2\left(\rho+m_{\sigma}^4\right)=2\mu\sqrt{1+\frac{H^2}{\mu^2}}\left(3-\beta+2\beta\frac{H^2}{\mu^2}\right).
\end{equation}
To match the standard cosmology on the brane with a vanishing cosmological constant in the limits $H^2/\mu^2\ll1$, one requires
\begin{equation}\label{eq:match}
\kappa_5^2m_{\sigma}^4=2\mu(3-\beta),\quad \mu\kappa_5^2=(1+\beta)\kappa_4^2,
\end{equation}
with which the modified FRW equation (\ref{eq:FRW1}) becomes
\begin{equation}\label{eq:FRW2}
(1+\beta)\frac{\rho}{\mu^2}+2(3-\beta)=2\sqrt{1+\frac{H^2}{\mu^2}}\left(3-\beta+2\beta\frac{H^2}{\mu^2}\right).
\end{equation}
The modified FRW equation (\ref{eq:FRW1}) can also be rewritten in terms of a dimensionless parameter $x$ as \cite{Lidsey:2003sj}
\begin{eqnarray}
H^2&=&\mu^2\left(\frac{1-\beta}{\beta}\cosh\left(\frac{2}{3}x\right)-\frac{1}{\beta}\right),\\
\rho&=&m_{\sigma}^4\left(\frac{m_{\alpha}^4}{m_{\sigma}^4}\sinh x-1\right),
\end{eqnarray}
where (we adopt the convention $\kappa_4^2=1$)
\begin{equation}
m_{\alpha}^4=\sqrt{\frac{2(1-\beta)^3}{\alpha\kappa_5^4}}=2\mu^2\sqrt{\frac{2(1-\beta)^3}{\beta(1+\beta)^2}}
\end{equation}
is a characteristic Gauss-Bonnet energy scale. Recalling that the Randall-Sundrum  energy scale now reads
\begin{equation}
m_{\sigma}^4=2\mu^2\left(\frac{3-\beta}{1+\beta}\right),
\end{equation}
one can classify the evolution of brane universe into three regimes:
the five-dimensional Gauss-Bonnet (GB) regime for $\rho\gg m_{\alpha}^4$ with modified FRW equation
\begin{equation}\label{eq:GB}
H^2\simeq\left(\frac{1+\beta}{4\beta}\mu\rho\right)^{\frac{2}{3}},
\end{equation}
the five-dimensional Randall-Sundrum (RS) regime for $m_{\alpha}^4\gg\rho\gg m_{\sigma}^4$ with modified FRW equation
\begin{equation}\label{eq:RS}
H^2\simeq\frac{1+\beta}{12(3-\beta)}\left(\frac{\rho}{\mu}\right)^2,
\end{equation}
and the five-dimensional general relativity (GR) regime for $m_{\sigma}^4\gg\rho$ with normal FRW equation
\begin{equation}\label{eq:GR}
H^2\simeq\frac{\rho}{3}.
\end{equation}

\begin{figure*}
  \includegraphics[width=8cm]{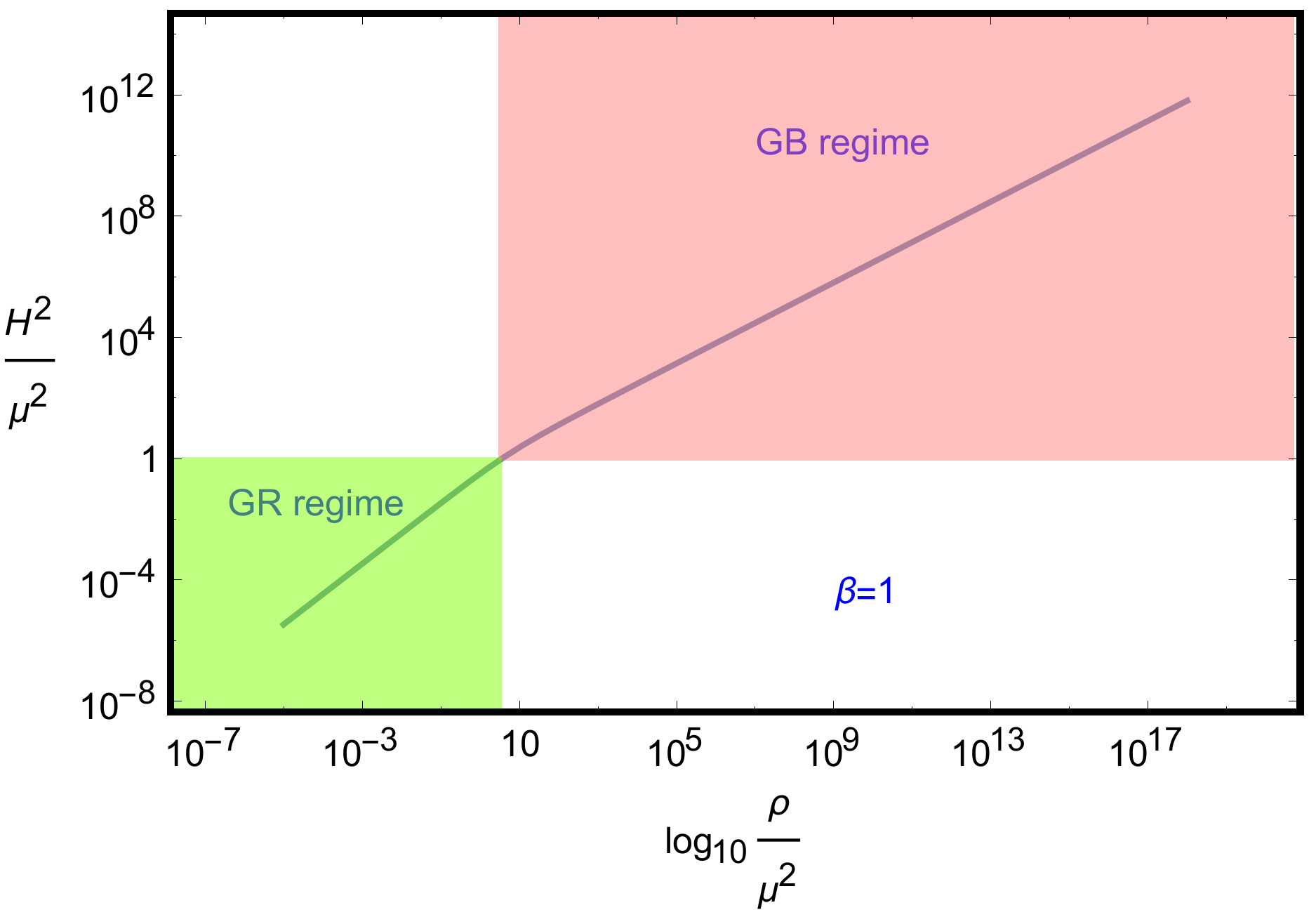}
  \includegraphics[width=8cm]{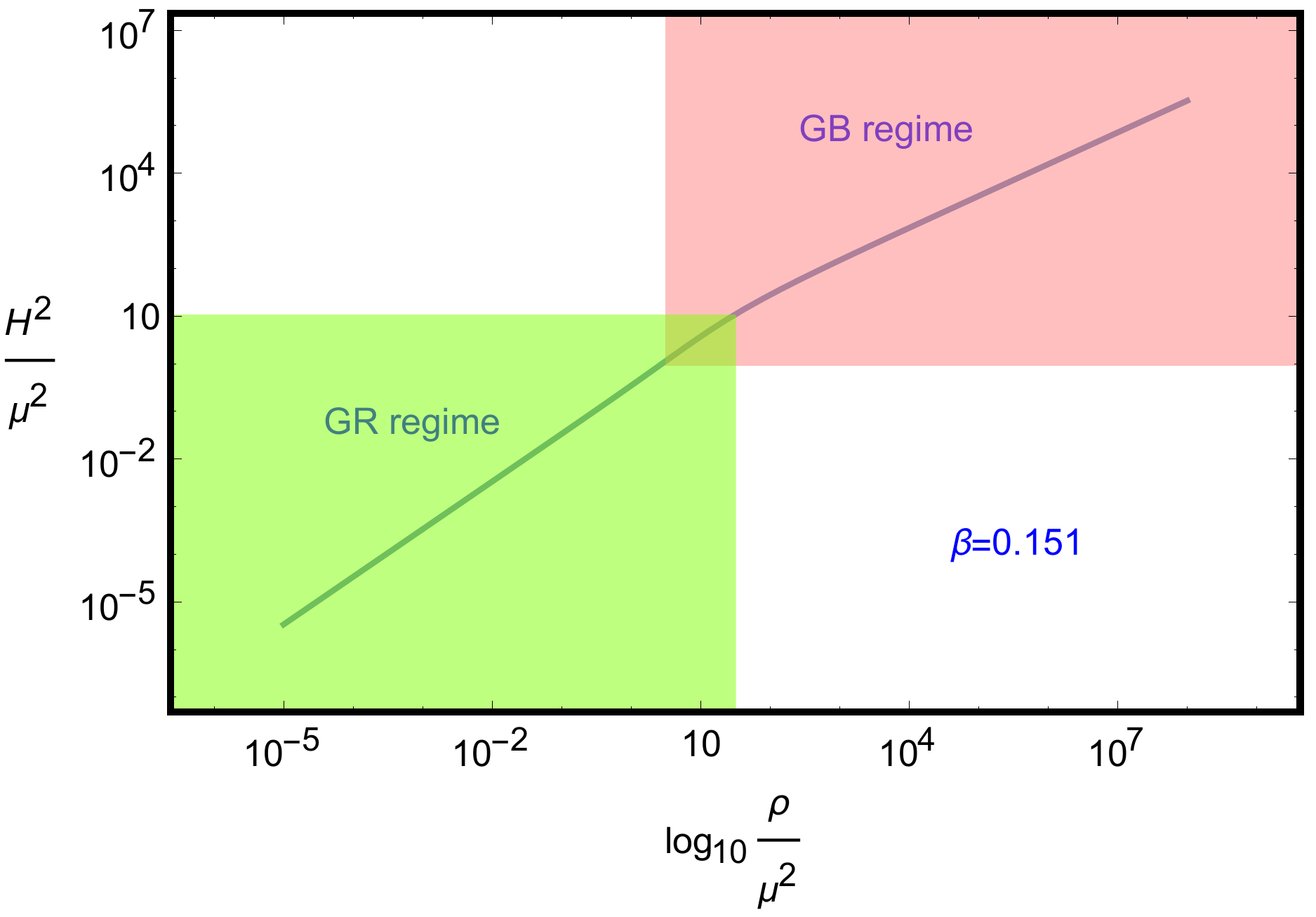}\\
  \includegraphics[width=8cm]{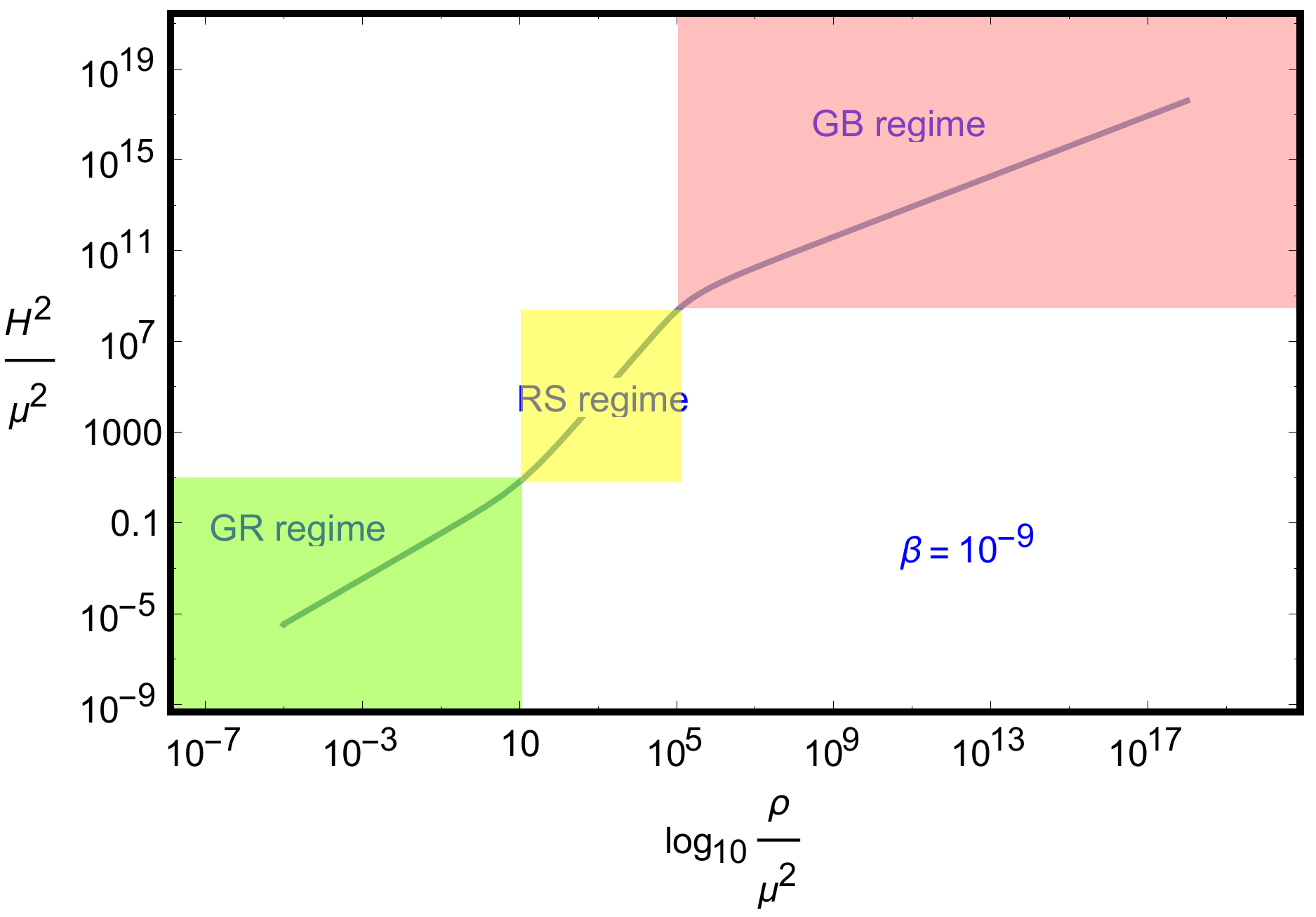}
  \includegraphics[width=8cm]{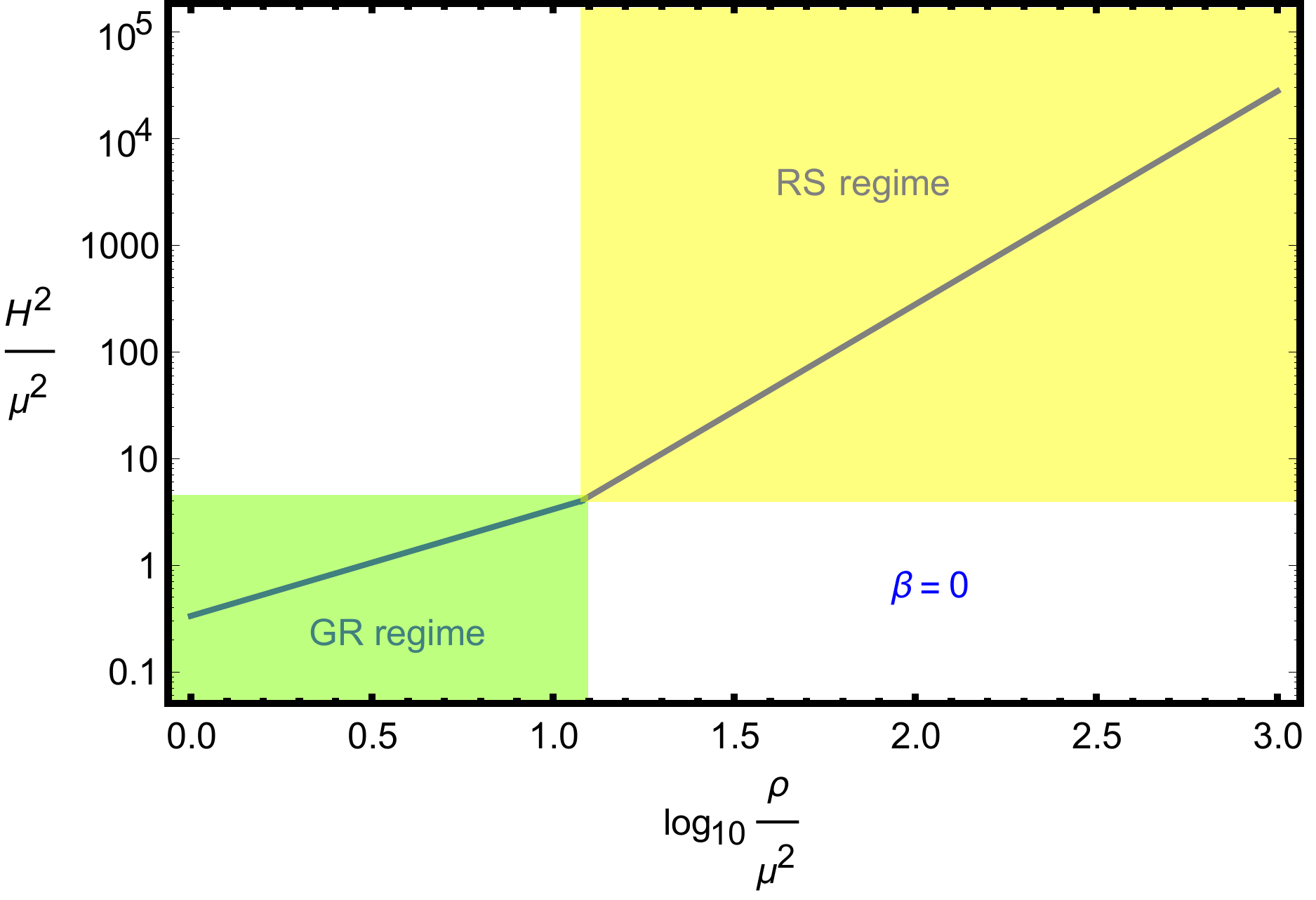}\\
  \caption{The evolution of brane universe with different choices of $\beta$. The vertical and horizontal axis describe the inflationary Hubble scale $H^2/\mu^2$ and energy density scale $\rho/\mu^2$ on the brane, respectively, for a given extra dimension scale $\mu$. With decreasing $\beta$ from $1$ to $0$, the GB regime is pushed toward to even higher energy scale and the RS regime grows slowly to finally dominate after its emergence when GB energy scale finally wins over the RS energy scale.}\label{fig:evolution}
\end{figure*}

In the high Hubble scale $H^2/\mu^2\gg1$, the modified FRW equation (\ref{eq:FRW2}) describes the GB regime (\ref{eq:GB}), while in the low Hubble scale $H^2/\mu^2\ll1$, the modified FRW equation (\ref{eq:FRW2}) describes the GR regime (\ref{eq:GR}). The RS regime emerges when the RS energy scale is smaller than the GB energy scale $m_{\sigma}<m_{\alpha}$, which is $\beta\lesssim0.151$. We will set $\beta\simeq0.151$ from now on to simplify the evolution of brane universe with GB regime followed immediately by GR regime. The full evolution of the brane universe is presented in Fig. \ref{fig:evolution}
with several typical choices of $\beta$.

\section{Higgs Inflation in the Gauss-Bonnet Braneworld}\label{sec:3}

We first review the Higgs inflation in four-dimensional Einstein gravity. The action in the Jordan frame is
\begin{equation}\label{eq:4DJordan}
S_4=\int\mathrm{d}^4x\sqrt{-g_4}\left(\frac{M_4^2}{2}\Omega^2R_4-\frac{1}{2}(\partial h)^2-V(h)\right),
\end{equation}
where $M_4^2\Omega^2=M^2+\xi h^2$ and $V(h)=\frac{\lambda}{4}(h^2-v^2)^2$. The four-dimensional Planck mass is recovered via $M_4^2=M^2+\xi v^2$ when the Higgs field is at its VEV $v=246$ GeV. As long as the nonminimal coupling $\xi\lll M_4^2/v^2\sim10^{32}$, one can safely approximate $M^2=M_4^2-\xi v^2\simeq M_4^2$. Therefore, after making conformal transformation
\begin{equation}
\Omega^2=\frac{\widetilde{g}_{\mu\nu}}{g_{\mu\nu}}=1+\frac{\xi h^2}{M_4^2},
\end{equation}
and field redefinition
\begin{equation}\label{eq:redefinition}
\left(\frac{\mathrm{d}\chi}{\mathrm{d}h}\right)^2
=\frac{1}{\Omega^2}+\frac{6M_4^2}{\Omega^2}\left(\frac{\mathrm{d}\Omega}{\mathrm{d}h}\right)^2,
\end{equation}
one has the action in the Einstein frame
\begin{equation}\label{eq:4DEinstein}
\widetilde{S}_4=\int\mathrm{d}^4x\sqrt{-\widetilde{g}}\left(\frac{M_4^2}{2}\widetilde{R}-\frac{1}{2}(\widetilde{\partial}\chi)^2-U(\chi)\right),
\end{equation}
where
\begin{equation}\label{eq:U}
U(\chi)=\frac{V(h(\chi))}{\Omega^4(h(\chi))}\simeq\frac{\lambda M_4^4}{4\xi^2}\left(1+e^{-\frac{2\chi}{\sqrt{6}M_4}}\right)^{-2}.
\end{equation}
Here we have used the large field solution $h=(M_4/\sqrt{\xi})\exp(\chi/\sqrt{6}M_4)$ of the field redefinition equation (\ref{eq:redefinition}) in large field limit $h\gg M_4/\sqrt{\xi}$.

Then we uplift the Ricci scalar curvature in (\ref{eq:4DEinstein}) as if it is reduced from the five-dimensional Gauss-Bonnet gravity when the extra dimension emerges at the high-energy scale. The action of our model then reads by choosing the matter field Lagrangian on the brane in (\ref{eq:action}) as the canonically normalized Higgs field,
\begin{equation}\label{eq:5DEinstein}
\mathcal{L}_{\mathrm{matter}}=-\frac{1}{2}(\partial\chi)^2-U(\chi).
\end{equation}
It is worth noting that, unlike previous works \cite{Farakos:2006sr}/\cite{Nozari:2008ny} where a bulk/brane scalar field nonminimally coupled to bulk/brane curvature in the Jordan frame, the canonically normalized Higgs field $\chi$ in the Einstein frame is minimally coupled to the five-dimensional Gauss-Bonnet gravity in our model. We argue that the action (\ref{eq:action}) with (\ref{eq:5DEinstein}) is actually a natural choice from effective field theory perspective. At the low-energy scale, the extra degrees of freedom due to the presence of the extra dimension should be integrated out and the physics should be well described by SM with a nonminimal coupling term. When the energy scale increases, the physical Higgs boson starts to decouple from SM, and it is the canonically normalized Higgs field that plays the role of inflaton. Therefore, the effect of the Gauss-Bonnet braneworld needs to be accounted for only when one goes to higher energy in the Einstein frame if extra dimension really exists. Thus we directly uplift the curvature term in (\ref{eq:4DEinstein}) as if it is reduced from the five-dimensional Gauss-Bonnet gravity at leading order.

Then the inflationary predictions \cite{Dufaux:2004qs} can be carried out directly just as those been done in \cite{Neupane:2014zxa}:
First, solving Hubble parameter from modified FRW equation (\ref{eq:FRW2}) by replacing $\rho$ with $U(\chi)$,
and calculating slow-roll parameters \cite{Okada:2014eva}
\begin{equation}
\epsilon(\chi)=\frac{U'(\chi)H'(\chi)}{3H(\chi)^3},\quad \eta(\chi)=\frac{U''(\chi)}{3H(\chi)^2},
\end{equation}
to find the endpoint $\chi_{\mathrm{end}}$ of inflation by solving $\max[\epsilon(\chi_{\mathrm{end}}),|\eta(\chi_{\mathrm{end}})|]=1$.
Second, solving $\chi_N$ and $\lambda/\xi^2$ from the combined equations \cite{Dufaux:2004qs}:
\begin{eqnarray}
N&=&\int_{\chi_{\mathrm{end}}}^{\chi_N}\mathrm{d}\chi\frac{3H(\chi)^2}{U'(\chi)},\\
A_s&=&\frac{9}{4\pi^2}\frac{H(\chi_N)^6}{U'(\chi_N)^2}\label{eq:As},
\end{eqnarray}
for given $e$-folding number $N$ and \emph{Planck} normalization $\ln(10^{10}A_s)=3.094$.
Third, with $\chi_N$ and $\lambda/\xi^2$ solved above, we can easily get the scalar spectral index
\begin{equation}
n_s=1-6\epsilon(\chi_N)+2\eta(\chi_N),
\end{equation}
its running
\begin{equation}
\alpha_s=\frac{U'(\chi_N)}{3H^2(\chi_N)}\left(6\epsilon'(\chi_N)-2\eta'(\chi_N)\right),
\end{equation}
and the tensor-to-scalar ratio $r=A_t/A_s$, where the amplitude of gravitational wave is given by \cite{Dufaux:2004qs}
\begin{equation}
A_t=\frac{2H(\chi_N)^2}{\pi^2}F^2\left(\frac{H(\chi_N)}{\mu}\right),
\end{equation}
with suppression factor
\begin{equation}\label{eq:suppression}
F(x)^2=\left(\sqrt{1+x^2}-\frac{1-\beta}{1+\beta}\,x^2\sinh^{-1}\frac{1}{x}\right)^{-1}.
\end{equation}
The pivot scale is chosen as $k=0.05\mathrm{Mpc}^{-1}$ and the $e$-folding number is taken in the range $N=50\sim60$.

\section{Tree-Level Results}\label{sec:4}

\begin{figure*}
  \includegraphics[width=8cm]{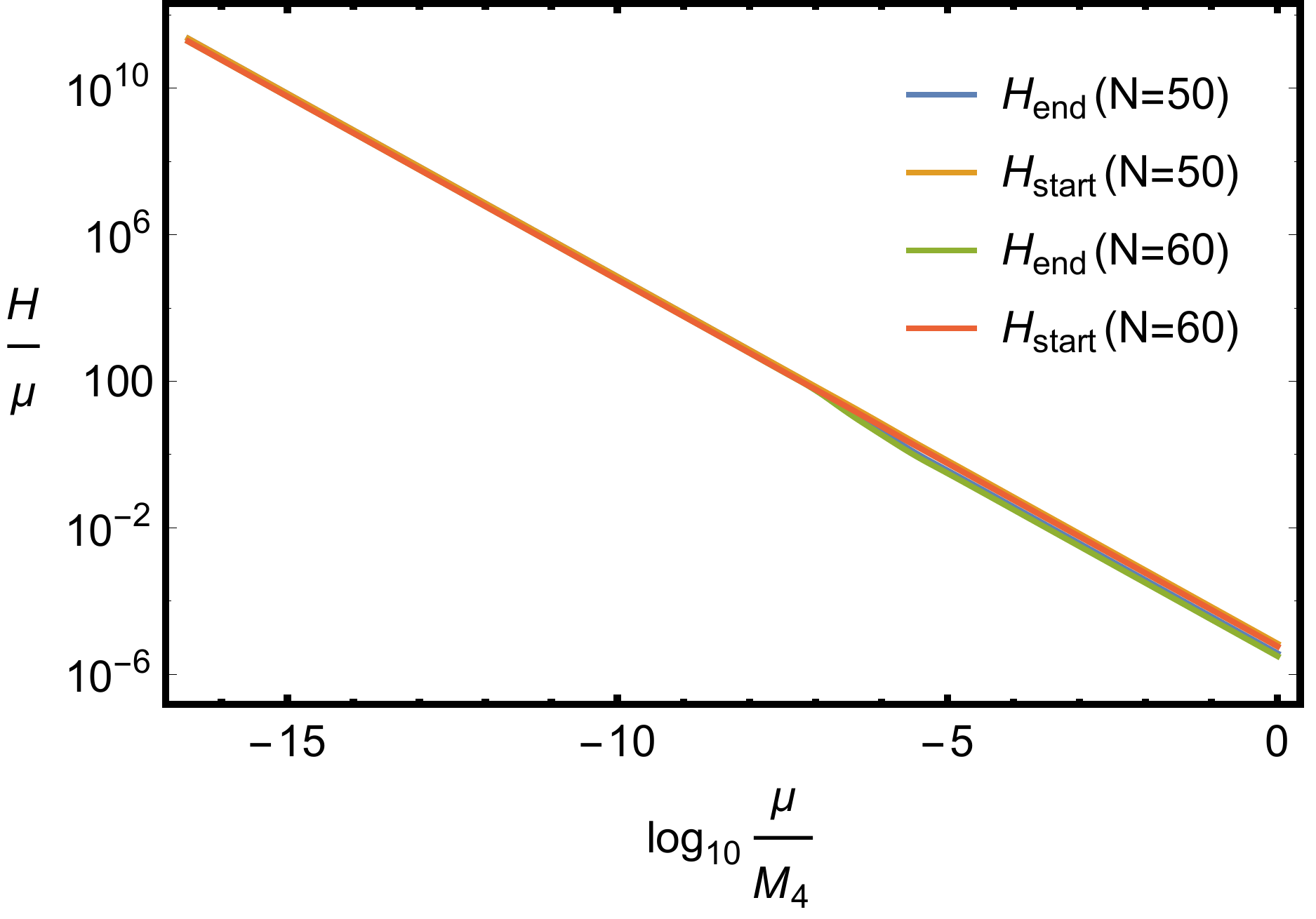}
  \includegraphics[width=8cm]{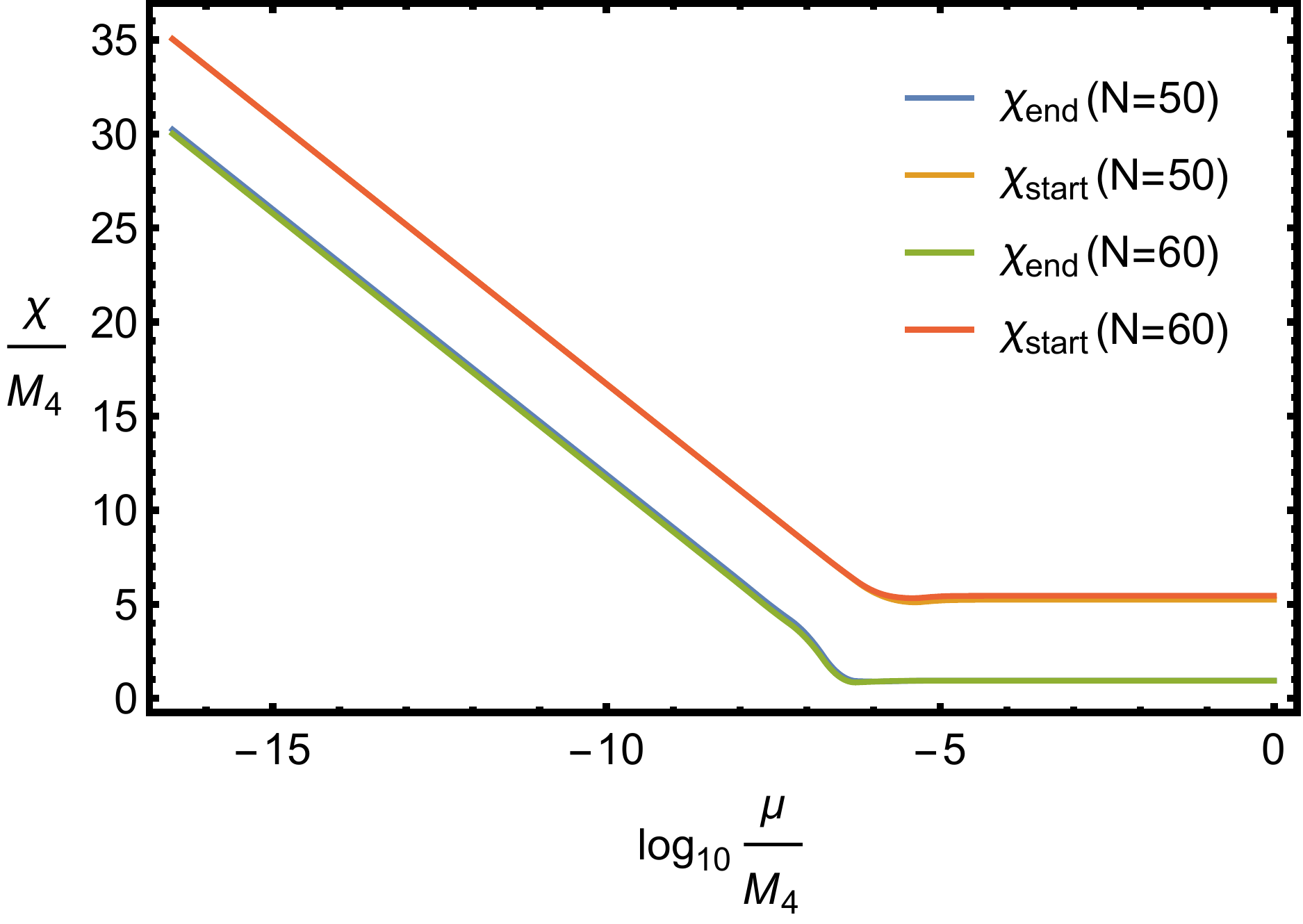}\\
  \includegraphics[width=8cm]{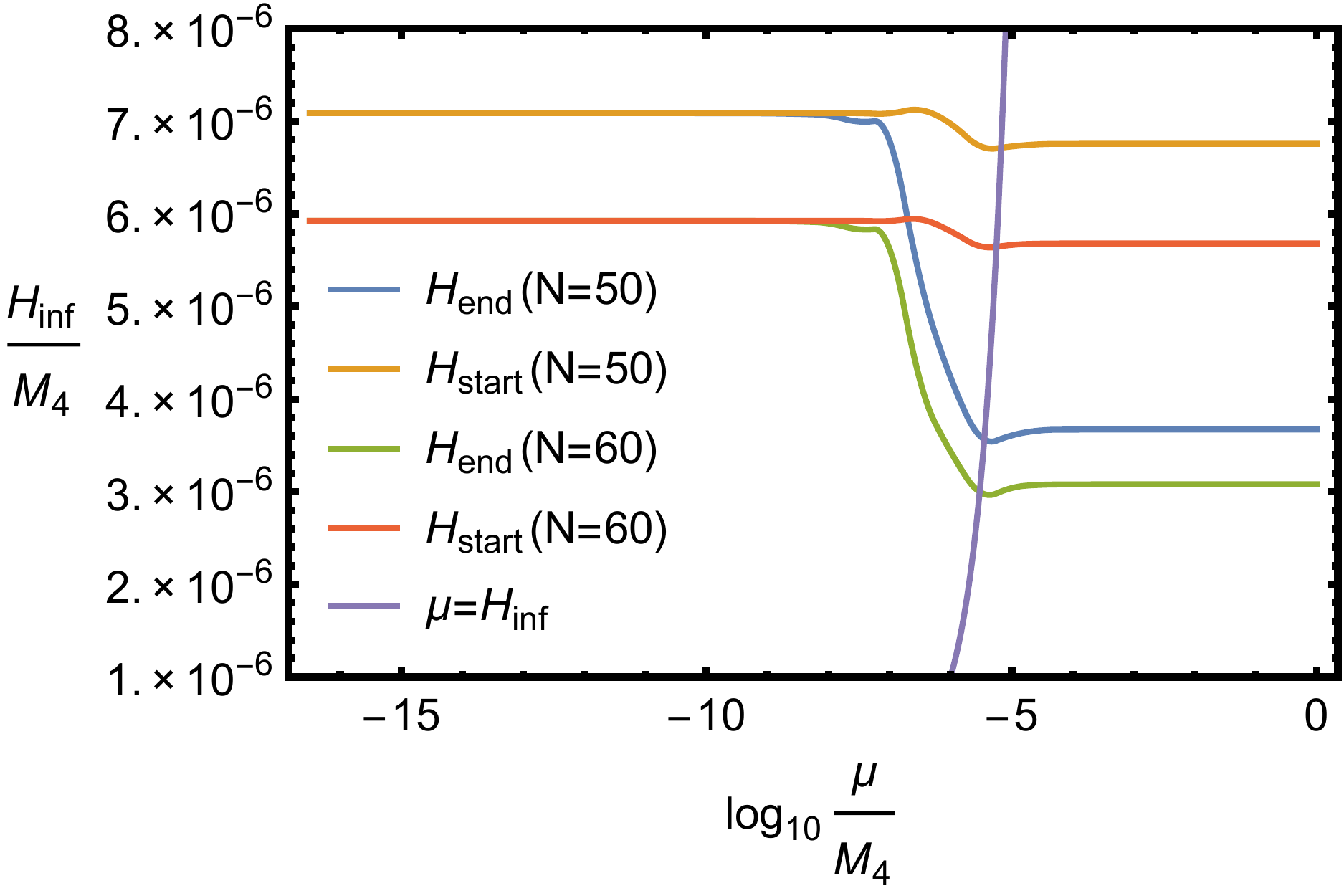}
  \includegraphics[width=8cm]{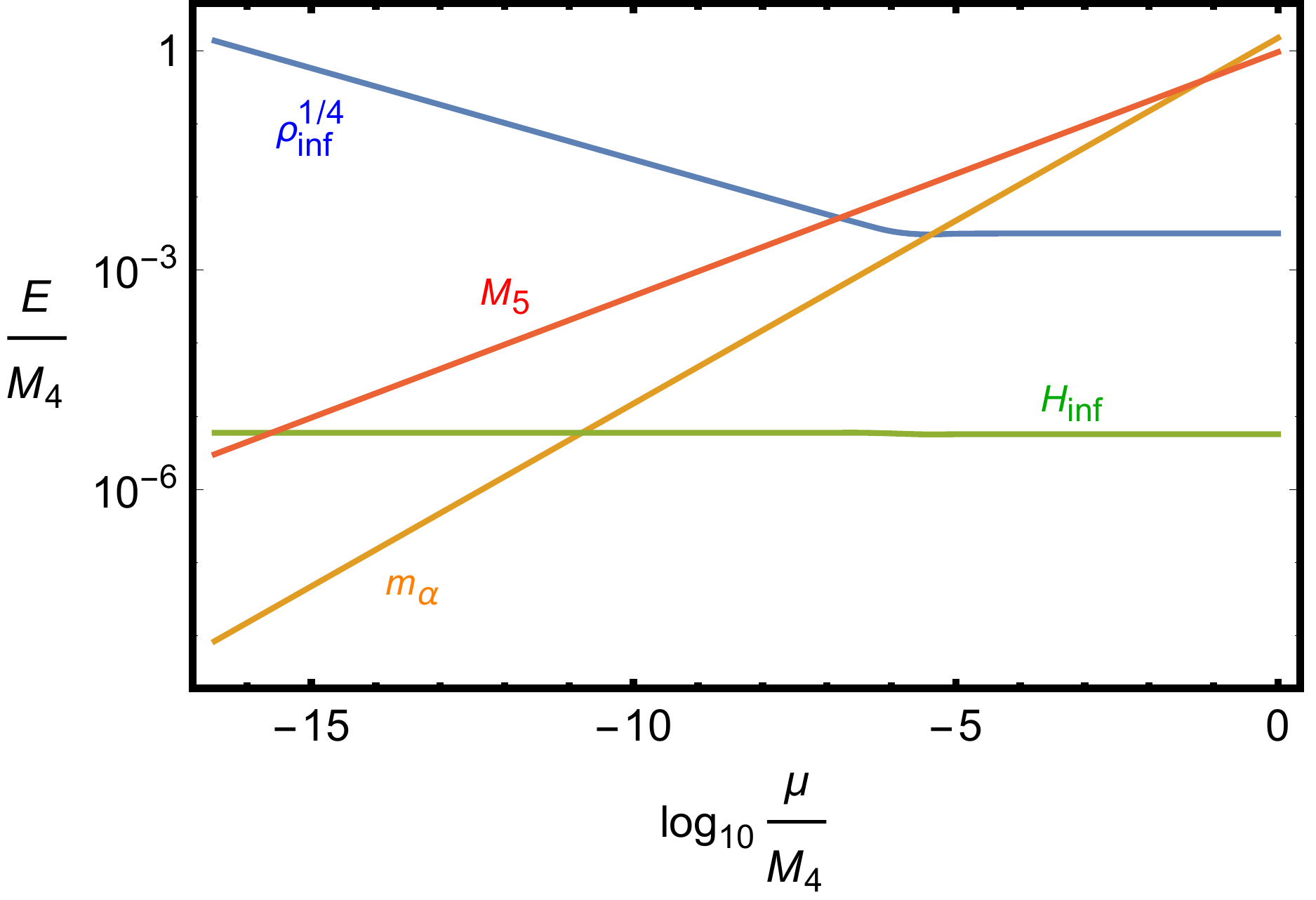}\\
  \caption{The characteristic features of Higgs inflation with respect to the energy scale $\mu$ of extra dimension in the Gauss-Bonnet braneworld. In the first panel, GR regime with $\frac{H}{\mu}\ll1$ is found for $\mu\gtrsim10^{12}\,\mathrm{GeV}$ and GB regime with $\frac{H}{\mu}\gg1$ is found for $\mu\lesssim10^{12}\,\mathrm{GeV}$. In the second panel, the field values at the start/end point of inflation remain constant in the GR regime but increase significantly in the GB regime. However, the field excursion of inflation is about $5M_4$ for all $\mu$. In the third panel, the inflationary Hubble scales are shown with respect to the extra dimension scale. In the left hand side of purple line where the inflationary Hubble scale equals to the extra dimension scale, the extra dimension scale appears below the inflationary Hubble scale. In the last panel, we summarize several typical energy scales with respect to the extra dimension scale.}\label{fig:inflation}
\end{figure*}

In Fig. \ref{fig:inflation},
we present various characteristic features of Higgs inflation with respect to the energy scale $\mu$ of the extra dimension in the Gauss-Bonnet braneworld.

The first result is that the extra dimension scale must be below the SM instability scale in order to have Higgs inflation in the GB regime. This can be seen from the first panel in Fig. \ref{fig:inflation}: the GB regime with $H/\mu\gg1$ is found for $\log_{10}(\mu/M_4)\lesssim-6$, which is roughly the energy scale $\mu\lesssim10^{12}\,\mathrm{GeV}$ where the Higgs quartic coupling would become negative. This result can also be derived formally as follows: in the GB regime one can use the modified FRW equation (\ref{eq:GB}) in the \emph{Planck} normalization (\ref{eq:As}) and find that
\begin{equation}
A_s=\frac{27\mu^2}{32\pi^2}\left(\frac{1+\beta}{4\beta}\right)^2\left(1+e^{\frac{2\chi_N}{\sqrt{6}M_4}}\right)^2.
\end{equation}
Since $\chi_N\gtrsim5M_4$ from the second panel of Fig. \ref{fig:inflation}, one immediately obtains
\begin{equation}\label{eq:upper}
\mu\lesssim3.4\times10^{12}\,\mathrm{GeV}.
\end{equation}

The second result is that the energy scale of the extra dimension lies below not only the SM instability scale but also the inflationary Hubble scale in the GB regime. This can be seen from the third panel of Fig. \ref{fig:inflation} where the purple line is obtained by $\mu=H$. The GB regime lies in the left hand side of purple line, where the extra dimension scale is less than the inflationary Hubble scale. Therefore, there might be KK modes excited during inflation and one should worry about whether these KK modes would spoil the flatness of inflationary potential. Fortunately, the spectrum of KK modes consists only of the massless four-dimensional graviton and a continuum of states with mass \cite{Dufaux:2004qs}
\begin{equation}
m>\frac{3}{2}H
\end{equation}
larger than the inflationary Hubble scale, which is too heavy to be excited during inflation in any real processes. And those scattering processes involving these heavy KK modes running in the loops are highly suppressed by their mass in the propagators. Therefore the flatness of inflationary potential is preserved. However, the GR regime lies in the right hand side of purple line, where the extra dimension scale is above the inflationary scale. Therefore, the extra dimension is invisible for the Higgs boson during inflation, thus Higgs inflation in the five-dimensional GR regime is effectively the same as four-dimensional Higgs inflation, and this is why we are not interested in Higgs inflation in GR regime.

The third result is that the Hubble scale remains almost unchanged during inflation in the GB regime. This can be seen from the third panel of Fig. \ref{fig:inflation} where $H_{\mathrm{start}}\thickapprox H_{\mathrm{end}}$ for a given $\mu$ in the GB regime. This can be understood as follows: On the one hand, we can see from the second panel of Fig. \ref{fig:inflation} that, the field values during inflation in the GB regime are deeper into the exponential plateaulike potential than in the GR regime; therefore, the potential change is rather small during inflation. On the other hand, the modified FRW equation $H^2\sim\rho^{2/3}$ in the GB regime suppresses the contributions to Hubble parameter from the potential changes. In general, the inflationary Hubble scale is
\begin{equation}\label{eq:Hubble}
H_{\mathrm{inf}}\simeq10^{13}\,\mathrm{GeV}.
\end{equation}

The fourth result is that the extra dimension scale must be above TeV scale. This can be seen from the last panel of Fig. \ref{fig:inflation} where we require that the inflationary Hubble scale is below the five-dimensional Planck scale $H_{\mathrm{inf}}\lesssim M_5$, namely,
\begin{equation}\label{eq:lower}
\mu\gtrsim1\,\mathrm{TeV}.
\end{equation}
It is consistent with the requirement that the energy density on the brane should be limited by the induced four-dimensional Planck scale $\rho_{\mathrm{inf}}\lesssim M_4^4$, which also leads to $\mu\gtrsim1\,\mathrm{TeV}$. For the extra dimension scale $\mu\sim1\,\mathrm{TeV}$, the five-dimensional Planck scale $M_5\sim1.7\times10^{13}\,\mathrm{GeV}$ is very closed to the inflationary Hubble scale (\ref{eq:Hubble}). With the upper bound (\ref{eq:upper}), the five-dimensional Planck scale can also be bounded from above, namely,
\begin{equation}
M_5\lesssim2.6\times10^{16}\,\mathrm{GeV}.
\end{equation}

\begin{figure*}
  \includegraphics[width=8cm]{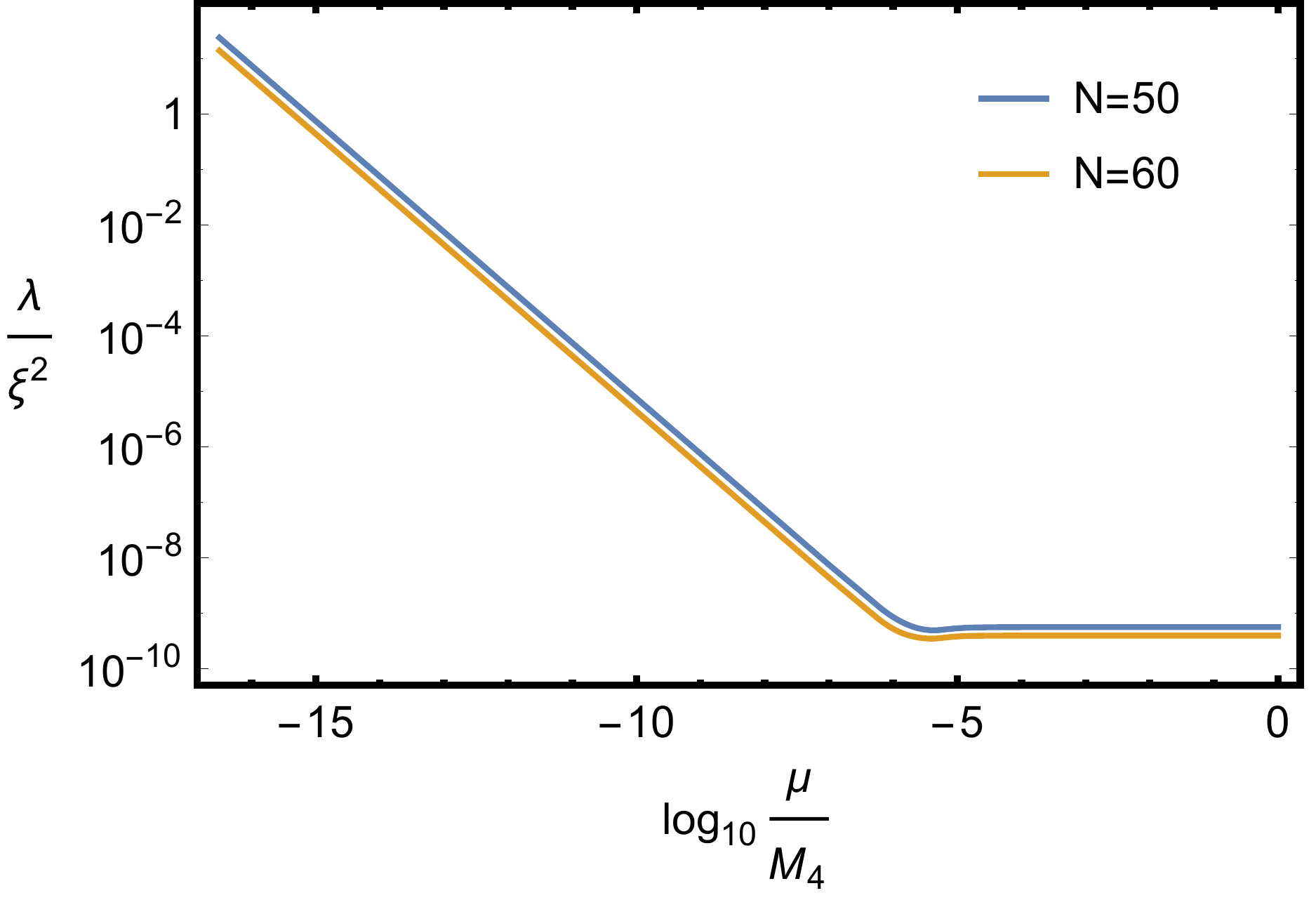}
  \includegraphics[width=8cm]{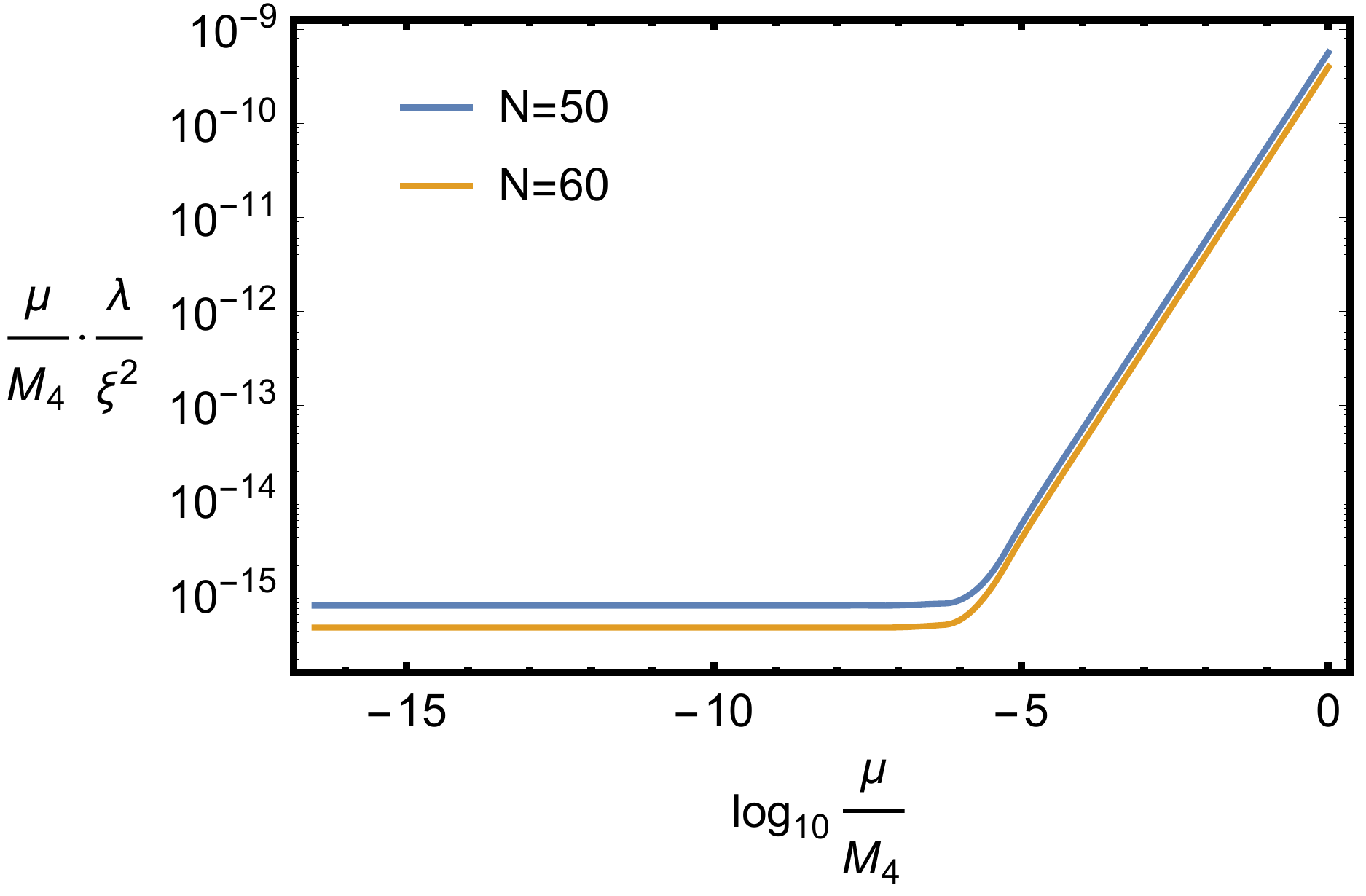}\\
  \caption{The prediction of combined parameter $\frac{\lambda}{\xi^2}$ for Higgs inflation with respect to the energy scale $\mu$ of extra dimension in the Gauss-Bonnet braneworld. For $\mu$ larger than $10^{12}\,\mathrm{GeV}$, $\frac{\lambda}{\xi^2}$ remains constant around $10^{-10}$ as in the four-dimensional case. For $\mu$ smaller than $10^{12}\,\mathrm{GeV}$, $\frac{\lambda}{\xi^2}$ increases significantly but the product $\frac{\mu}{M_4}\frac{\lambda}{\xi^2}$ remains constant around $10^{-16}$. For $\mu$ around TeV scale, $\frac{\lambda}{\xi^2}$ is of order $\mathcal{O}(0.1)$, which can lead to a naturally small $\xi\sim\mathcal{O}(1)$ for $\lambda\sim\mathcal{O}(0.1)$.}\label{fig:parameter}
\end{figure*}

In Fig. \ref{fig:parameter}
we present the combined parameter $\lambda/\xi^2$ for Higgs inflation with respect to the extra dimension scale $\mu$ in the Gauss-Bonnet braneworld. In the GR regime with modified FRW equation (\ref{eq:GR}), the combined parameter
\begin{equation}
\frac{\lambda}{\xi^2}\simeq\frac{3}{4}\frac{H_{\mathrm{inf}}^2}{M_4^2}\simeq10^{-10},
\end{equation}
remains almost the same as in the four-dimensional case. In the GB regime, $\lambda/\xi^2$ surprisingly increases many orders of magnitudes with decreasing the extra dimension scale. However, with modified FRW equation (\ref{eq:GB}), the product
\begin{equation}
\frac{\mu}{M_4}\frac{\lambda}{\xi^2}\simeq\frac{H_{\mathrm{inf}}^3}{M_4^3}\frac{16\beta}{1+\beta}\simeq10^{-16},
\end{equation}
remains constant with decreasing the extra dimension scale in GB regime. In the case with the extra dimension scale near the TeV scale, which is of interest in experiments, $\lambda/\xi^2$ can achieve order $\mathcal{O}(0.1)$, which can be made by
\begin{equation}
\xi\sim\mathcal{O}(1),\quad\lambda\sim\mathcal{O}(0.1).
\end{equation}
This naturally solves the unitarity problem without fine-tuning $\lambda$ and violating \emph{Planck} 2015 $\mathrm{TT,TE,EE+lowP}$ bound $r\lesssim0.1$ as shown shortly below.

Next we explain why we are only interested in Higgs inflation in the GB regime instead of the RS regime. In the GB regime, the requirement that the inflationary Hubble scale is below the five-dimensional Planck scale,
\begin{equation}
\frac{1+\beta}{16\beta}\frac{\lambda}{\xi^2}\mu M_4^2\simeq H_{\mathrm{inf}}^3\lesssim M_5^3=\frac{\mu M_4^2}{1+\beta},
\end{equation}
namely,
\begin{equation}
\frac{\lambda}{\xi^2}\lesssim\frac{16\beta}{(1+\beta)^2}\simeq1.82,
\end{equation}
is satisfied as long as the extra dimension scale being just larger than the TeV scale. However, if we allow $\beta$ to take other values near $0$, then Higgs inflation can also take place in the RS regime with modified FRW equation (\ref{eq:RS}), which can give
\begin{equation}
\frac{\lambda}{\xi^2}\simeq\frac{\mu}{M_4}\frac{H_{\mathrm{inf}}}{M_4}\sqrt{\frac{192(3-\beta)}{1+\beta}}.
\end{equation}
To naturally solve the unitarity problem, one needs $\lambda/\xi^2$ being of order $\mathcal{O}(0.1)$, which requires both the extra dimension scale and inflationary Hubble scale being extremely near the four-dimensional Planck scale. Therefore, Higgs inflation in the RS regime is less interesting than in the GB regime from experimental point of view.

\begin{figure*}
  \includegraphics[width=8cm]{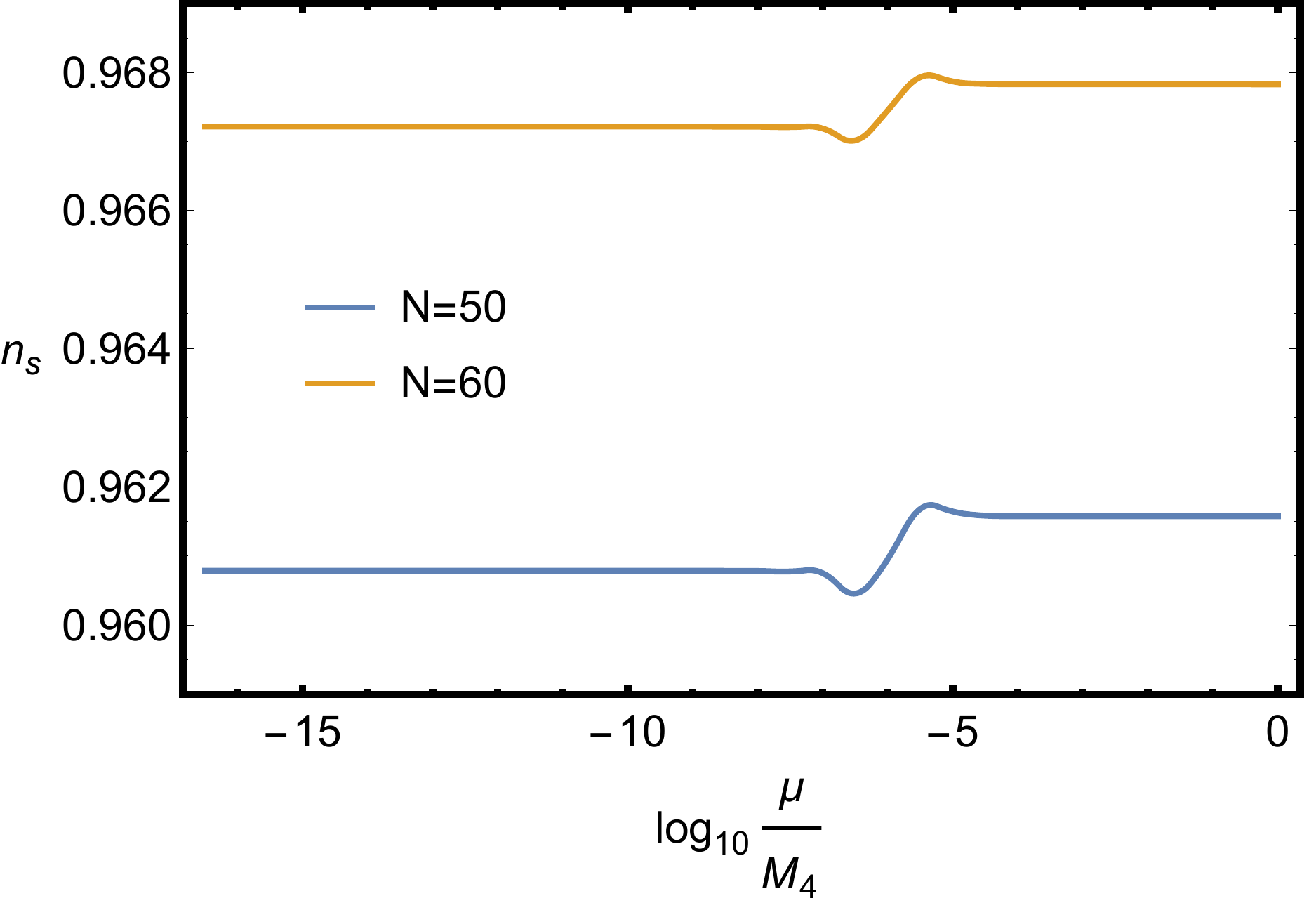}
  \includegraphics[width=8cm]{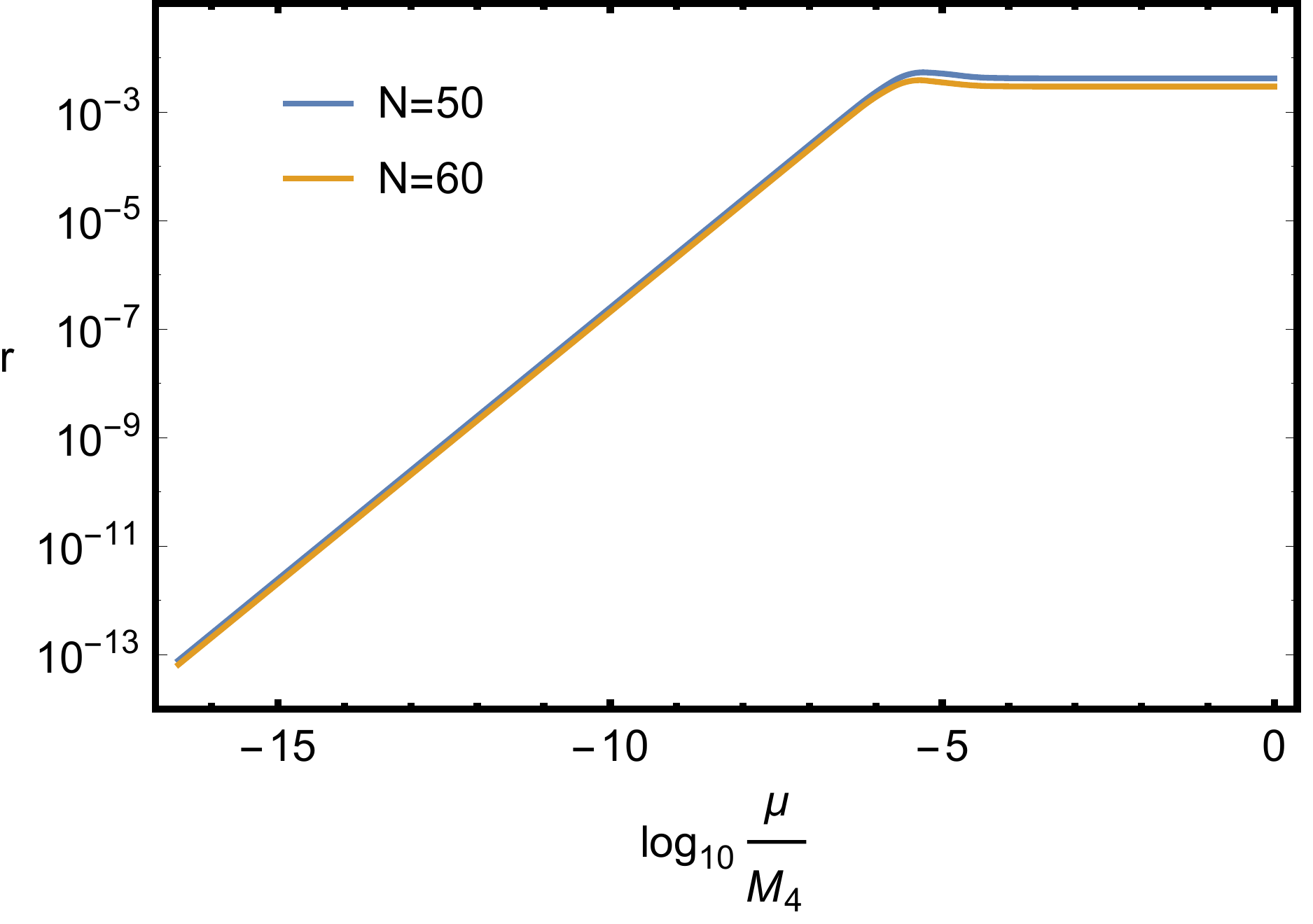}\\
  \includegraphics[width=8cm]{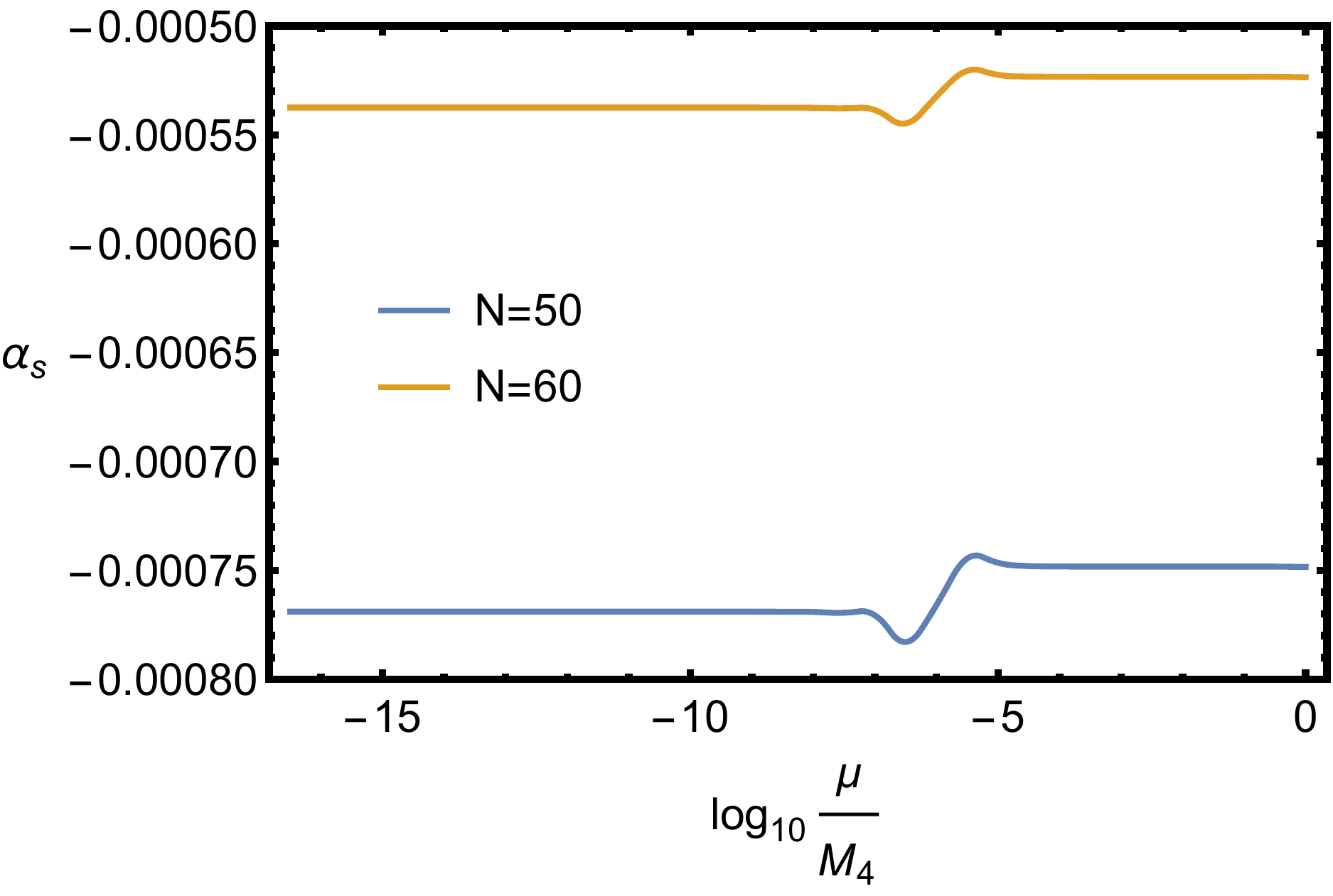}
  \includegraphics[width=8cm]{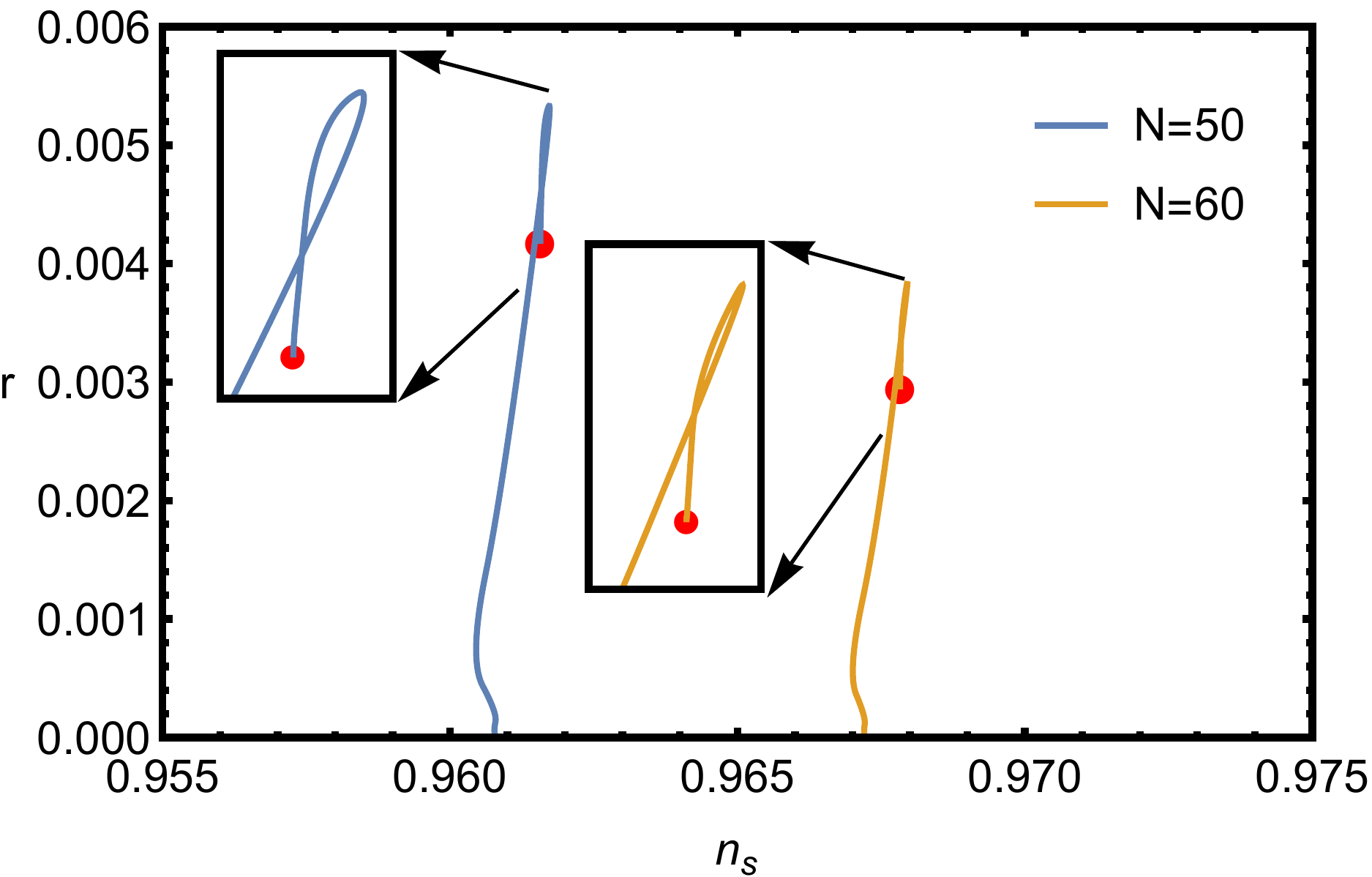}\\
  \caption{Inflationary predictions for Higgs inflation in the Gauss-Bonnet braneworld . The first three panels show the predictions of scalar spectral index $n_s$, tensor-to-scalar ratio $r$, and the running of scalar spectral index $\alpha_s$ with respect to the energy scale $\mu$ of extra dimension. With decreasing $\mu$, $r$ drops significantly while $n_s$ and $\alpha_s$ remain almost the same as in the four-dimensional case. In the last panel, the inflationary predictions of $n_s$ and $r$ are showed in the $n_s\!-\!r$ plane, where the red points represent the case of four-dimensional Higgs inflation.}\label{fig:prediction}
\end{figure*}

In Fig. \ref{fig:prediction}
we present inflationary predictions of $n_s$, $r$, and $\alpha_s$ for Higgs inflation in the Gauss-Bonnet braneworld. Both $0.960\lesssim n_s\lesssim0.968$ and $-0.0008\lesssim\alpha_s\lesssim-0.0005$ are stable against the change of the extra dimension scale. Only in the GR regime $r\sim10^{-3}$ as in the four-dimensional case.  In the GB regime $r$ drops significantly with the decreasing extra dimension scale. Unlike critial Higgs inflation \cite{Hamada:2013mya,Cook:2014dga,Hamada:2014iga,Bezrukov:2014bra,Hamada:2014wna} in four dimensions where $r\gtrsim0.1$ for the $\xi\sim\mathcal{O}(1)$, our $r$ can be as small as $10^{-12}$ safely inside \emph{Planck} 2015 $\mathrm{TT,TE,EE+lowP}$ bound $r\lesssim0.1$ for extra dimension scale around TeV scale. Recall that the field excursion during inflation is roughly $5M_4$ for all extra dimension scale, it explicitly evades the usual argument of the Lyth bound that a super-Planckian field excursion during inflation corresponds to an observable tensor-to-scalar ratio. The Lyth bound is modified in the presence of extra dimension in our model
\begin{equation}
\int_{\chi_{\mathrm{end}}}^{\chi_N}\mathrm{d}\chi=\int_0^N\mathrm{d}N\sqrt{\frac{r}{8F^2}},
\end{equation}
where the suppression factor (\ref{eq:suppression}) is $F^2(\frac{H}{\mu})\thickapprox1$ for the GR regime and $F^2(\frac{H}{\mu})\thickapprox\frac{1+\beta}{2\beta}\frac{\mu}{H}$ for the GB regime, namely,
\begin{equation}
\frac{\Delta\chi}{M_4}=\int_0^N\left(\frac{\mathrm{d}N}{55}\right)\left(\frac{H}{M_4}\right)^{\frac{1}{2}}\left(\frac{\mu}{M_4}\right)^{-\frac{1}{2}}\left(\frac{r}{0.01}\right)^{\frac{1}{2}}.
\end{equation}
With decreasing the extra dimension scale, the tensor-to-scalar ration can be actually dragged down to the unobservable level even if the field excursion during inflation is super-Planckian.

\section{Going Beyond Tree-level Analysis}\label{sec:5}

In the last section we see that for Higgs inflation to take place in the GB regime, the extra dimension scale is below the inflationary Hubble scale. Thanks to the fact that the masses of extra KK modes are larger than the inflationary Hubble scale, the flatness of inflationary potential is preserved since these heavy KK modes certainly cannot be excited at external legs and any contributions from these heavy KK modes running in the loop are suppressed by their masses in the propagators. When going beyond tree-level analysis of the renormalization-group(RG)-improved effective potential, we can actually follow the methods \cite{DeSimone:2008ei,Bezrukov:2008ej,Bezrukov:2009db,Allison:2013uaa,Salvio:2013rja} developed in the four-dimensional Higgs inflation. The net effect of adding the extra dimension to Higgs inflation is the change of normalization condition (\ref{eq:As}) for scalar spectrum amplitude due to the modified FRW equation (\ref{eq:FRW2}) at the background level. We present below the procedures to carry out the predictions of $\xi$, $\lambda$, $n_s$, $\alpha_s$ and $r$ at inflationary scale with respect to the top quark mass for a given Higgs mass at electroweak scale and an extra dimension scale.

First, the initial conditions at $\bar{\mu}=m_t$ for the $\overline{\mathrm{MS}}$ SM couplings are taken from \cite{Buttazzo:2013uya}, which are repeated here for convenience:
\begin{eqnarray}
g'(m_t)&=&0.3587,\\
g(m_t)&=&0.6483,\\
\nonumber g_s(m_t)&=&1.1666+0.00314\frac{\alpha_s(m_Z)-0.1184}{0.0007}\\
                  &-&0.00046\left(\frac{m_t}{\mathrm{GeV}}-173.35\right),\\
\nonumber y_t(m_t)&=&0.93697+0.00550\left(\frac{m_t}{\mathrm{GeV}}-173.35\right)\\
\label{eq:th1}    &-&0.00042\frac{\alpha_s(m_Z)-0.1184}{0.0007}\pm0.00050_{\mathrm{th}},\\
\nonumber\lambda(m_t)&=&0.12710+0.00206\left(\frac{m_h}{\mathrm{GeV}}-125.66\right)\\
\label{eq:th2}       &-&0.00004\left(\frac{m_t}{\mathrm{GeV}}-173.35\right)\pm0.00030_{\mathrm{th}}.
\end{eqnarray}

Second, the three-loop RG equations for SM couplings, 3-loop RG equation for the Higgs anomalous dimension $\gamma=\mathrm{d}\ln h/\mathrm{d}\ln\bar{\mu}$ and two-loop RG equation for the nonminimal coupling are used in our analysis from the Appendix of Ref. \cite{Allison:2013uaa}. We omit here the complete expressions for these RG equations. However, it is worth noting that the $s$ factor \cite{DeSimone:2008ei,Bezrukov:2009db}
\begin{equation}
s(h)=\frac{1+\xi h^2/M_4^2}{1+(1+6\xi)\xi h^2/M_4^2}
\end{equation}
insertions will be important in our case of small $\xi$, unlike the four-dimensional Higgs inflation with large $\xi$ which renders a chiral electroweak theory at the high-energy scale. Solve these RG equations with above initial conditions and input parameters within corresponding uncertainties \cite{Buttazzo:2013uya}
\begin{eqnarray}
m_h&=&(125.66\pm0.34)\,\mathrm{GeV},\\
m_t&=&(173.36\pm0.65\pm0.3)\,\mathrm{GeV},\\
\alpha_s(m_Z)&=&0.1184\pm0.0007,
\end{eqnarray}
we can get the running couplings and anomalous dimesion as functions of renormalization scale $\bar{\mu}=m_te^t$.

Third, we do not include the two-loop radiative corrections in our effective potential since the tree-level potential
\begin{equation}
U_0(\chi)=\frac{\lambda h^4}{4\Omega^4},
\end{equation}
and one-loop Coleman-Weinberg potential
\begin{align}
U_1(\chi)=&\frac{1}{16\pi^2}\left[\frac{3}{2}M_W^4\left(\ln\frac{M_W^2}{\bar{\mu}^2}-\frac{5}{6}\right)\right.\\
\nonumber &+\frac{3}{4}M_Z^4\left(\ln\frac{M_Z^2}{\bar{\mu}^2}-\frac{5}{6}\right)
           -3M_t^4\left(\ln\frac{M_t^2}{\bar{\mu}^2}-\frac{3}{2}\right)\\
\nonumber &+\left.\frac{1}{4}M_h^4\left(\ln\frac{M_h^2}{\bar{\mu}^2}-\frac{3}{2}\right)
           +\frac{3}{4}M_G^4\left(\ln\frac{M_G^2}{\bar{\mu}^2}-\frac{3}{2}\right)\right],
\end{align}
where
\begin{align}
\nonumber &M_W^2=\frac{g^2h^2}{4\Omega^2},\,M_Z^2=\frac{(g^2+g'^2)h^2}{4\Omega^2},\,M_t^2=\frac{y_t^2h^2}{2\Omega^2},\\
&M_h^2=\frac{3s\lambda h^2}{\Omega^4}\left(\frac{1-\xi h^2/M_4^2}{1+\xi h^2/M_4^2}\right),\,M_G^2=\frac{\lambda h^2}{\Omega^4}
\end{align}
are sufficient for our purpose. Note that we are working in the prescription I, where quantum corrections are computed in the Einstein frame. After the following replacements \cite{Bezrukov:2008ej,Bezrukov:2009db},
\begin{eqnarray}
h&\rightarrow&Z(\bar{\mu})h,\\
\bar{\mu}&\rightarrow&\frac{h}{\Omega(h)},
\end{eqnarray}
where
\begin{eqnarray}
Z(\bar{\mu})&=&\exp\left(\int_{m_t}^{\bar{\mu}}\gamma(\bar{\mu}')\mathrm{d}\ln\bar{\mu}'\right),\\
\Omega(h)&=&\sqrt{1+\frac{\xi_0h^2}{M_4^2}},
\end{eqnarray}
we obtain the RG improved effective potential $U_{\mathrm{eff}}(\chi(h))=U_0(\chi(h))+U_1(\chi(h))$ where $\chi(h)$ is the solution of field redefinition (\ref{eq:redefinition}).

Finally, the initial value $\xi_0=\xi(m_t)$ can be fixed by matching the \emph{Planck} normalization (\ref{eq:As}) with $U(\chi)$ replaced by $U_{\mathrm{eff}}(\chi)$. To be more specific, for given input parameters of $m_h$, $m_t$, $\alpha_s(m_Z)$, $\xi_0$ and extra dimension scale $\mu$, we compute the initial conditions and then solves RG equations to get the running coupling and anomalous dimension.  We then compute the effective potential and scalar spectrum amplitude. Repeat above procedure by choosing different $\xi_0$ until the \emph{Planck} normalization is fulfilled. Once the $\xi_0$ is determined, we can follow the above procedures once again to obtain $\xi_{\mathrm{inf}}=\xi(h_N)$, $\lambda_{\mathrm{inf}}=\lambda(h_N)$ at $e$-folding number $N=60$. We can also obtain the corresponding values for $n_s$ and $r$ with $U(\chi)$ replaced by $U_{\mathrm{eff}}(\chi)$.

\begin{figure*}
  \includegraphics[width=8cm]{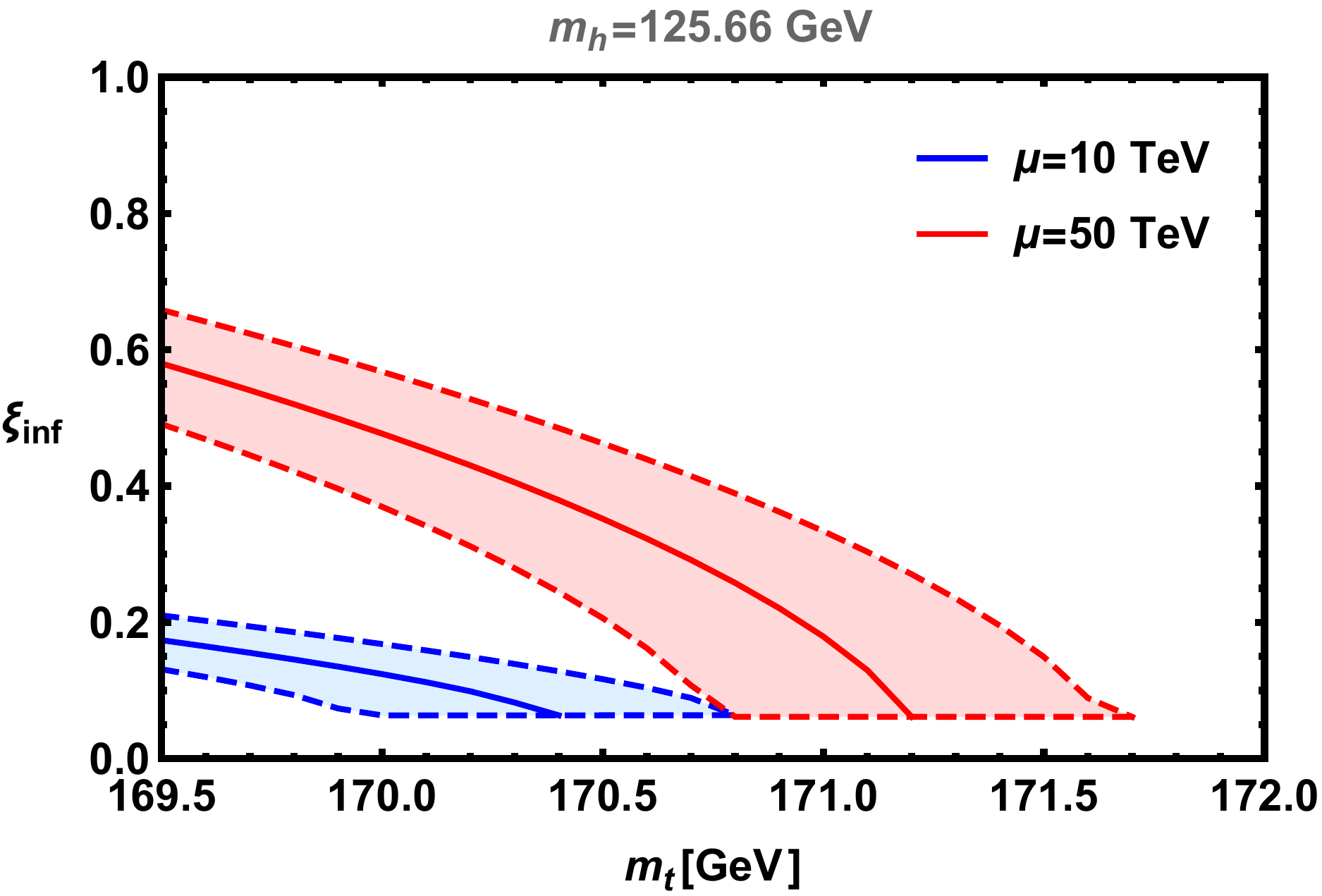}
  \includegraphics[width=8cm]{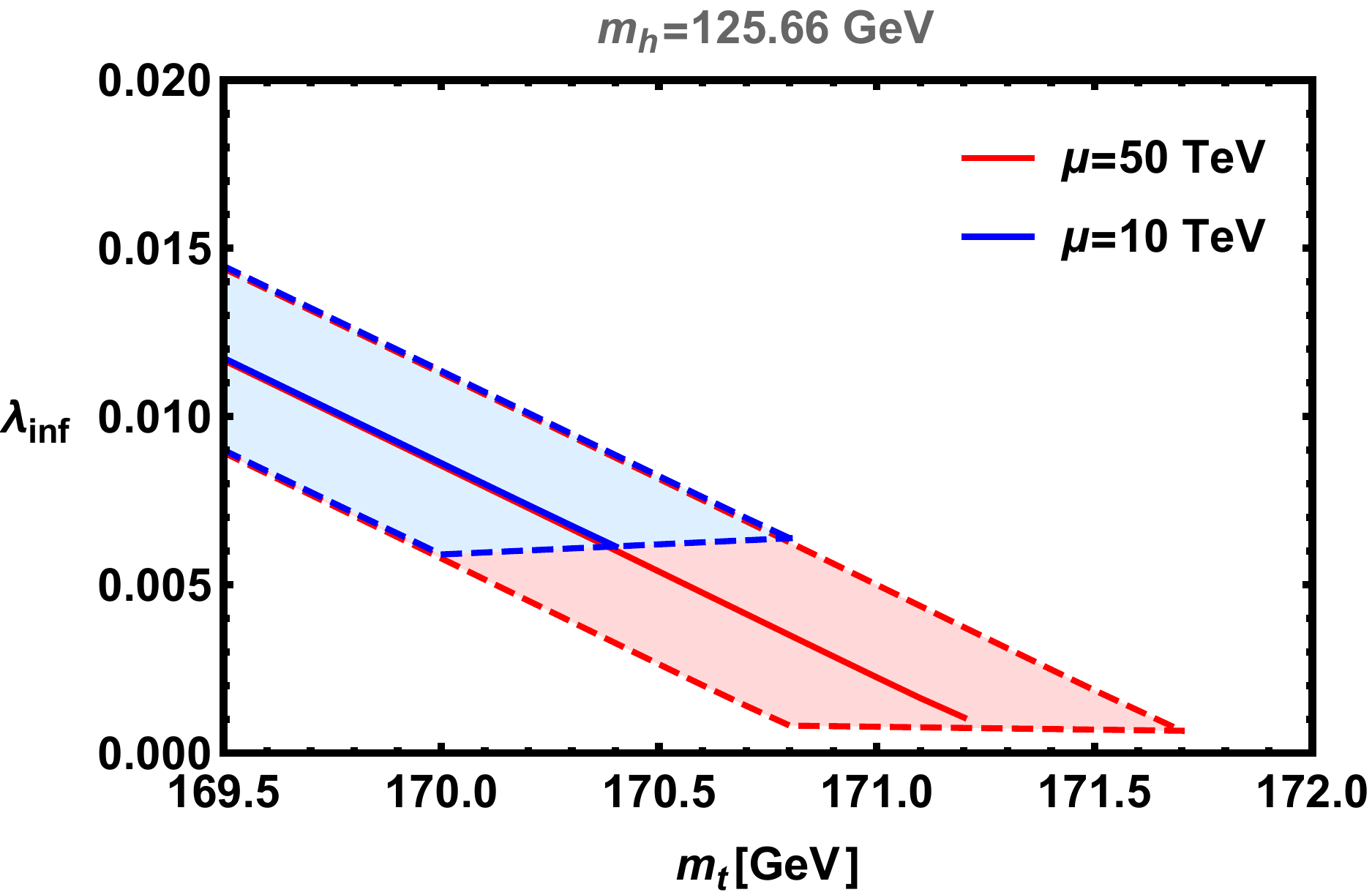}\\
  \includegraphics[width=8cm]{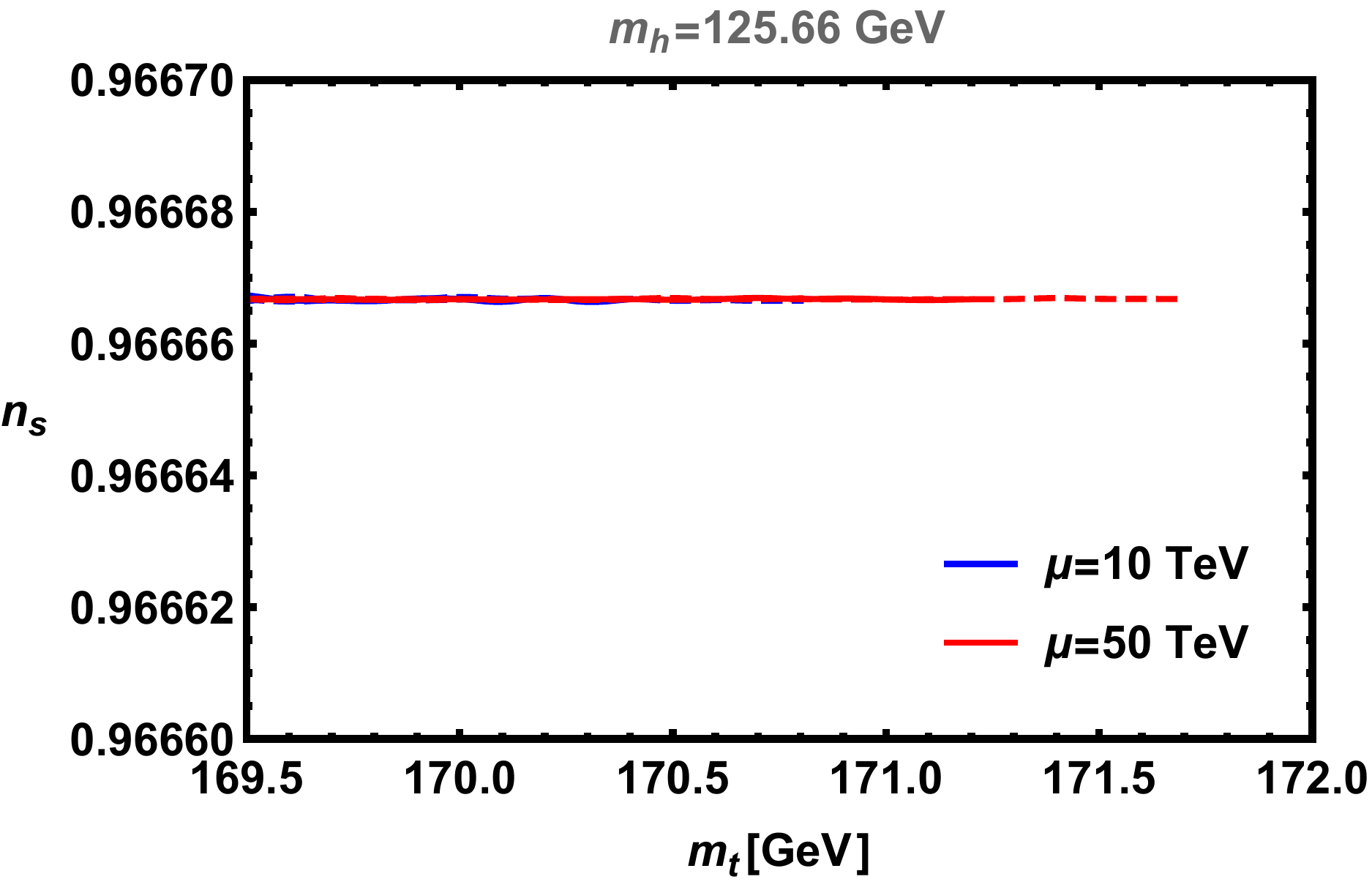}
  \includegraphics[width=8cm]{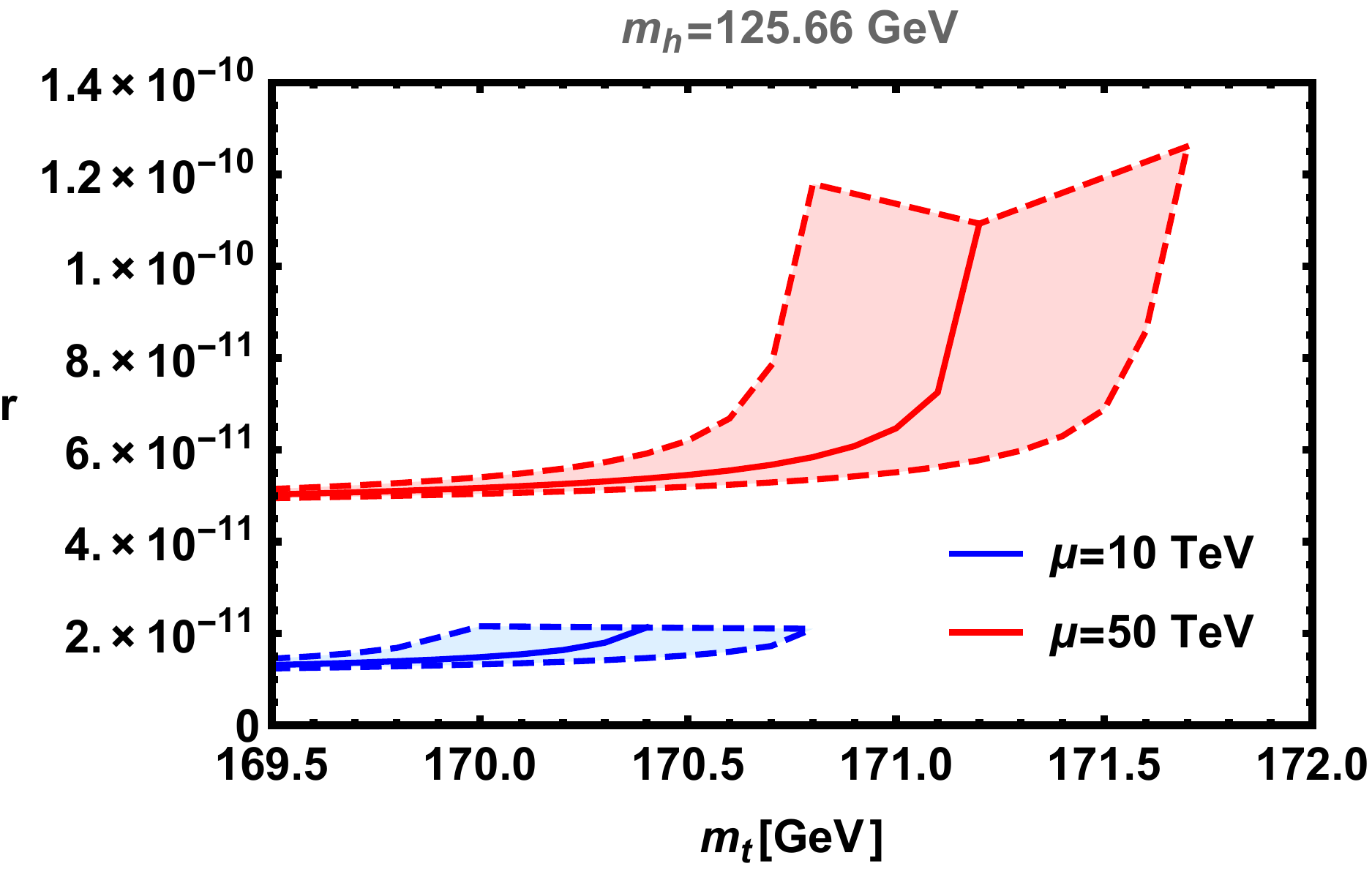}\\
  \caption{Numerical results of nonminimal coupling $\xi_{\mathrm{inf}}$, Higgs quartic coupling $\lambda_{\mathrm{inf}}$, scalar spectral index $n_s$ and tensor-to-scalar ratio $r$ during inflation with respect to the top quark mass assuming the Higgs mass $m_h=125.66\,\mathrm{GeV}$. The extra dimension scale is $10$ TeV for the blue region and $50$ TeV for the red region respectively, where the $1\sigma$ uncertainties are mainly from strong coupling $\alpha_s$ along with other theoretical uncertainties from the threshold corrections.}\label{fig:loop}
\end{figure*}

In Fig. \ref{fig:loop}, with input Higgs mass $m_h=125.66$ GeV,
we present the numerical results of the nonminimal coupling $\xi_{\mathrm{inf}}$, Higgs quartic coupling $\lambda_{\mathrm{inf}}$, scalar spectral index $n_s$ and tensor-to-scalar ratio $r$ during inflation with respect to the top quark mass. The extra dimension scale is $10$ TeV for the blue region and $50$ TeV for the red region respectively, where the $1\sigma$ uncertainties are mainly from strong coupling $\alpha_s$ along with other theoretical uncertainties from the threshold corrections (\ref{eq:th1}) and (\ref{eq:th2}). During inflation, the nonminimal coupling $\xi_{\mathrm{inf}}\simeq\mathcal{O}(0.1)$ and the Higgs quartic coupling $\lambda_{\mathrm{inf}}\simeq\mathcal{O}(0.01)$. The inflationary predictions of $n_s$ and $r$ remain the same as the tree-level results.

We also find the following upper bound for the top quark mass as the function of the Higgs mass and strong coupling and other theoretical uncertainties when the extra dimension scale is fixed at $\mu=50$ TeV,
\begin{eqnarray}
\nonumber \frac{m_t}{\mathrm{GeV}}&<&171.179+0.4816\left(\frac{m_h}{\mathrm{GeV}}-125.66\right)\\
&&+0.283\left(\frac{\alpha_s(m_Z)-0.1184}{0.0007}\right)\pm0.162.
\end{eqnarray}

\begin{figure*}
  \includegraphics[width=15cm]{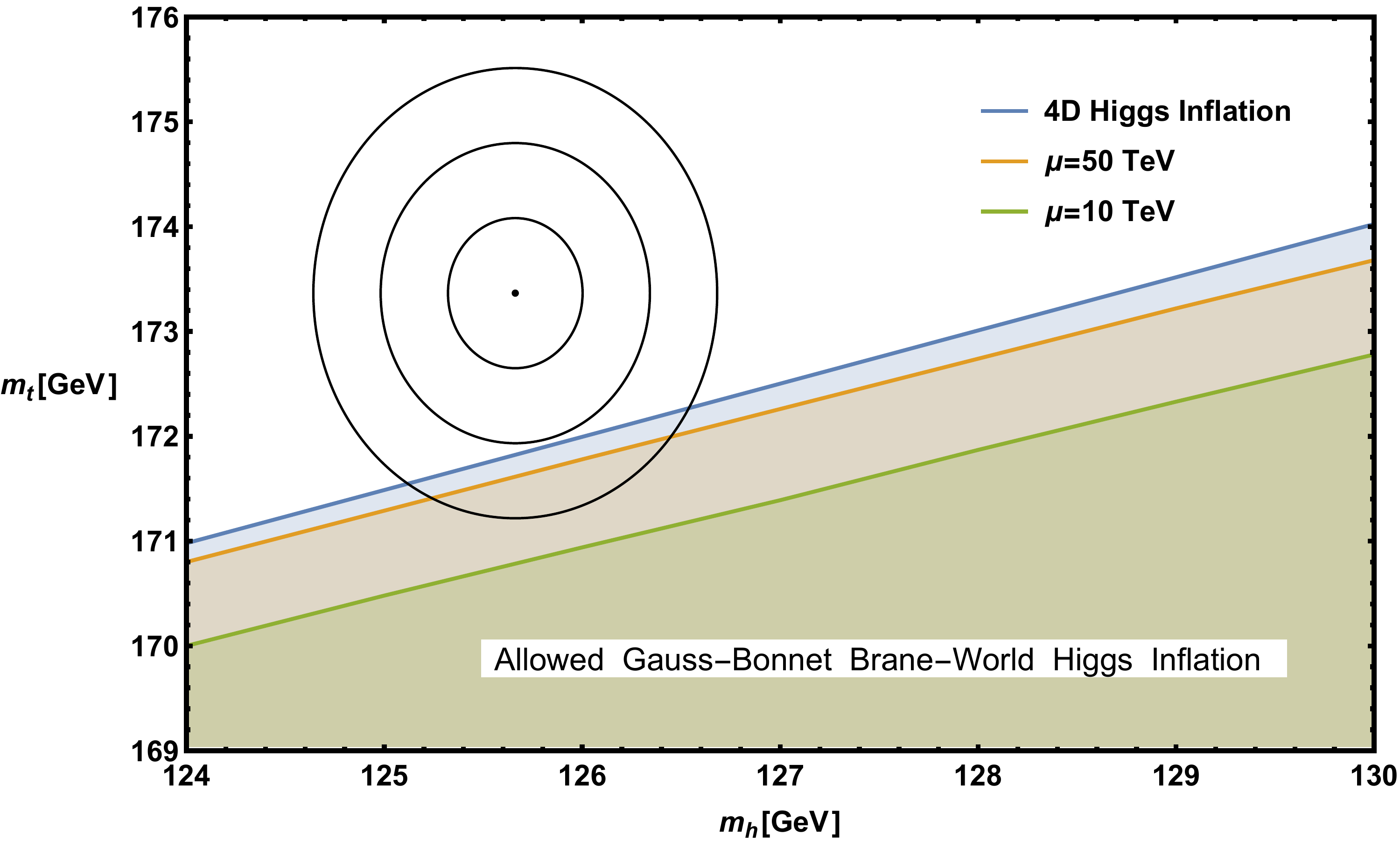}\\
  \caption{The upper bound of top quark mass $m_t$ as a function of Higgs mass $m_h$ where we have taken into account the $1\sigma$ uncertainties from strong coupling $\alpha_s$ along with other theoretical uncertainties from the threshold corrections. As comparisons, we also provide the upper bound set by four-dimensional Higgs inflation and the experimental constraint on $m_t$ and $m_h$.}\label{fig:lhc}
\end{figure*}

In Fig. \ref{fig:lhc},
we present the allowed region in $m_h-m_t$ plane for our model, where the $1\sigma$ uncertainties from the strong coupling $\alpha_s$ along with other theoretical uncertainties from the threshold corrections having been properly accounted for. We also provide the upper bound set by four-dimensional Higgs inflation and the experimental constraint on $m_t$ and $m_h$ for comparisons. We find that with increasing the energy scale of the extra dimension, the upper bound of the top quark mass of our model will approach toward to those in four-dimensional Higgs inflation. Therefore the stability problem still insists as in the four-dimensional Higgs inflation although the unitarity problem is indeed solved.

Fortunately the first nonzero KK mode appears above the inflationary scale although the extra dimension scale can be as low as TeV scale, therefore adding extra dimension only changes the background dynamics of the universe without jeopardizing the low-energy particle physics. To solve the stability problem, one could follow in the same spirit of stabilizing the SM effective potential and further generalizations of our model should be considered. For example, it was found in \cite{Gogoladze:2008ak,He:2012ub,Okada:2015zfa} that a TeV scale type III seesaw mechanism can simultaneously account for the neutrino oscillations and stabilize the SM effective potential without introducing any additional scalar fields. Therefore, the Higgs inflation in the Gauss-Bonnet braneworld can be self-consistent up to the inflationary Hubble scale which is free of the unitarity problem and stability problem. Although shift symmetry may still be needed to preserve the flatness of the potential above inflationary scale, which is a common problem shared by many other inflation models, Higgs inflation in the Gauss-Bonnet braneworld might be more convenient to be embedded into underlying UV theories.

\section{Conclusions}\label{sec:6}

In this paper, we realize Higgs inflation in the five-dimensional Gauss-Bonnet braneworld scenario. We find that, for Higgs inflation to take place in the GB regime, the extra dimension scale must be in the range between the TeV scale and the instability scale of SM. Furthermore, the intriguing improvement of many orders of magnitude for $\lambda/\xi^2$ with decreasing the extra dimension scale comes as a nice surprise. For the extra dimension scale around the experimentally interesting TeV scale, the nonminimal coupling can be made of order $\xi\sim\mathcal{O}(1)$ for the Higgs quartic coupling $\lambda\sim\mathcal{O}(0.1)$. The predicted scalar spectral index $0.960\lesssim n_s\lesssim0.968$ and its running $-0.0008\lesssim\alpha_s\lesssim-0.0005$ are well inside the \emph{Planck} 2015 constraints on inflation. Unlike the critical Higgs inflation, the tensor-to-scalar ratio $r\sim10^{-12}$ is safely inside \emph{Planck} 2015 $\mathrm{TT,TE,EE+lowP}$ bound $r\lesssim0.1$. We also investigate the inflationary predictions beyond tree-level analysis and find that the predictions remain almost the same as the tree-level results. However, to avoid the stability problem, one has to follow in the same spirit of stabilizing the SM effective potential by using, for instance, the TeV scale type III seesaw mechanism.

\begin{acknowledgments}
We would like to thank Mikhail Shaposhnikov for helpful discussions during the International Conference on Gravitation and Cosmology and the Fourth Galileo-Xu Guangqi Meeting at the Institute of Theoretical Physics, Chinese Academy of Sciences, Beijing, China.
R.G.C. is supported by the Strategic Priority Research Program of the Chinese Academy of Sciences, Grant No.XDB09000000.
Z.K.G. is supported by the National Natural Science Foundation of China, Grants No.11175225 and No.11335012.
\end{acknowledgments}

\bibliographystyle{apsrev4-1}
\bibliography{ref}

\end{document}